\documentclass[11pt]{statsoc}\usepackage[]{graphicx}\usepackage[]{color}
\makeatletter
\def\maxwidth{ %
  \ifdim\Gin@nat@width>\linewidth
    \linewidth
  \else
    \Gin@nat@width
  \fi
}
\makeatother

\definecolor{fgcolor}{rgb}{0.345, 0.345, 0.345}

\usepackage{framed}
\makeatletter
 {\par\unskip\endMakeFramed%
 \at@end@of@kframe}
\makeatother

\definecolor{shadecolor}{rgb}{.97, .97, .97}
\definecolor{messagecolor}{rgb}{0, 0, 0}
\definecolor{warningcolor}{rgb}{1, 0, 1}
\definecolor{errorcolor}{rgb}{1, 0, 0}

\usepackage{alltt}

\usepackage[a4paper]{geometry}
\usepackage{vmargin} 
\usepackage{float,booktabs}
\usepackage{graphicx}
\usepackage{hyperref}
\usepackage{numprint}

\usepackage{amsmath,amsfonts}
\usepackage{natbib}
 \usepackage{color-edits}
 \usepackage{url}

\usepackage{mhequ}
\newcommand{\be}{\begin{equs}}
\newcommand{\ee}{\end{equs}}
\newcommand{\mc}[1]{\mathcal{#1}}
\newcommand{\bb}[1]{\mathbb{#1}}

\DeclareMathOperator{\Bern}{Bernoulli}
\DeclareMathOperator{\diag}{diag}
\DeclareMathOperator{\Dir}{Dirichlet}
\DeclareMathOperator{\Cat}{Categorical}
\DeclareMathOperator{\var}{var}
\DeclareMathOperator{\train}{train}

\title[Closer than they appear]{Closer than they appear: A Bayesian perspective on individual-level heterogeneity in risk assessment}

\author{Kristian Lum}
\address{Department of Computer and Information Science, University of Pennsylvania, Philadelphia, USA.}

\author{David B. Dunson}
\address{Department of Statistical Science, Duke University, Durham, USA.}

\author[Lum {\it et al.}]{James Johndrow}
\address{Wharton School, University of Pennsylvania, Philadelphia, USA.}
\email{johndrow@wharton.upenn.edu}
\IfFileExists{upquote.sty}{\usepackage{upquote}}{}
\begin{document}

\allowdisplaybreaks
\maketitle

\begin{abstract}
Risk assessment instruments are used across the criminal justice system to estimate the probability of some future behavior given covariates. The estimated probabilities are then used in making decisions at the individual level. In the past, there has been controversy about whether the probabilities derived from group-level calculations can meaningfully be applied to individuals. Using Bayesian hierarchical models applied to a large longitudinal dataset from the court system in the state of Kentucky, we analyze variation in individual-level probabilities of failing to appear for court and the extent to which it is captured by covariates. We find that individuals within the same risk group vary widely in their probability of the outcome. In practice, this means that allocating individuals to risk groups based on standard approaches to risk assessment, in large part, results in creating distinctions among individuals who are not meaningfully different in terms of their likelihood of the outcome. This is because uncertainty about the probability that any particular individual will fail to appear is large relative to the difference in average probabilities among any reasonable set of risk groups. 
\end{abstract}


\section{Introduction}

Actuarial risk assessment instruments (RAIs) are used throughout the criminal justice system to issue data-driven estimates of the likelihood that a contextually relevant outcome will occur. RAIs have been used in criminal justice for decades \citep{harcourt2008against, solow2019institutional}, and for pre-trial decision making since at least the Vera Institute of Justice's ``point scale" was developed in the 1960s \citep{paulsen1966pre}. In recent years they have played a key role in reform efforts designed to reduce pre-trial detention and eliminate cash bail in the United States. For example, a central component of New Jersey's suite of reforms put in place to overhaul pre-trial decision making was the use of a popular RAI, the Public Safety Assessment (PSA) \citep{mdrc-newjersey}. Similarly, California's Senate Bill 10--- legislation passed in 2018 with the intention of reforming California's pre-trial system--- mandated the use of risk-based pre-trial decision making across the state \citep{solow2019institutional}. 

Despite their adjacency to more progressive criminal justice reform efforts, the fairness of these models has been questioned. For example, in an address to the National Association of Criminal Defense Lawyers that was otherwise optimistic about the uses of RAIs, former United States Attorney General Eric Holder, cautioned that ``[RAIs] may exacerbate unwarranted and unjust disparities that are already far too common in our criminal justice system" \citep{holder2014}. Legal scholars and civil society groups have raised concerns about the civil rights implications of their use \citep{robinson2019civil}. Perhaps most famously, \cite{angwin2016machine} of the investigative journalism organization, ProPublica, published a report claiming that one particular RAI was ``biased against blacks," largely based on the finding that the false positive rate of the model was higher for Black people than for white people. This (dis)parity-based notion of fairness has been the focus of much of the technical literature on fairness in risk assessment over the last several years \citep{flores2016false, dieterich2016compas, chouldechova2017fair}.

 Other technical concerns related to the fairness of RAIs center on statistical uncertainty. For example, \cite{hart2007precision} argued that standard measures of statistical uncertainty for RAIs pertained to groups, not individuals. Moreover,  
 they argued that the existence of large statistical uncertainty at the {\it individual-level} makes such models ill-suited to inform decisions about {\it individuals}.
 Extending this idea, we believe one related standard of ``fairness" is that individuals who are not statistically distinguishable by the model should not be (or rarely be) treated differently on the basis of the model's predictions. Assessing whether RAIs satisfy these standards of fairness requires fitting models explicitly designed to account for individual-specific probabilities and their associated uncertainty intervals. Without repeated observations of the same individuals over time, this is not possible \citep{imrey2015commentary}. Unfortunately, perhaps due to lack of availability of data of this type, to our knowledge there has been no work {\it in this domain} that uses appropriate statistical models to meaningfully estimate individual probabilities and associated uncertainty about them, leaving the question of whether existing RAIs meet these standards of fairness largely unknown. 
 
In this paper, we attempt to fill this gap. 
Using data on over \numprint{460000} arrests from mid-2014 to the end of 2018 in the state of Kentucky, we build a Bayesian hierarchical model with individual-level random effects. The outcome we model is failure to appear (FTA) for any court appointment prior to the conclusion of the case. We condition on covariates available at decision time, including the factors used in a popular pre-trial risk assessment instrument, Arnold Ventures' PSA. Taking advantage of our large, longitudinal dataset on FTA in which there are many individuals who appear on multiple occasions, we are able to estimate this model, and, importantly, learn the distribution of individual probabilities across the population. The main assumption of our model is that, while probabilities of failing to appear do vary across individuals in a way that is not explained by covariates, they only vary across time for a given individual $i$ in a manner explained by covariates.

 We later verify that our model probabilities are well-calibrated with respect to several strategies for creation of risk groups. Analysis of the distribution of individual-level probabilities of FTA across the population, in combination with data available to us, gives rise to some surprising conclusions about the extent to which individuals in the data are distinguishable with respect to their individual-level propensities toward the FTA outcome, even when we have been able to observe the individual's behavior on several previous occasions. To wit, we find that most individuals are effectively indistinguishable from one another. We conclude by explaining that the result is not an inevitable consequence of our model construction, and that, were the population-level distribution of individual probabilities more concentrated, or the available covariates much stronger predictors of the outcome, we could have reached the opposite conclusion ---i.e. that many individuals can be distinguished on the basis of their individual propensity toward the outcome. 

\section{What are risk assessment instruments}

At their core, actuarial RAIs are statistical models that correlate a set of covariates (often, factors that are grounded in criminological theory) to the relevant criminal justice-related outcome \citep{desmarais2019pre}. In this paper, we focus on prediction of FTA. However, the methods and concepts we use could be applied to risk assessment for virtually any outcome variable at any stage of the criminal justice process. In the canonical setting, actuarial RAIs estimate the probability, $P$, of a binary outcome, $Y$, conditional on observable covariates, $X \in \mc X$--- a framework common to many prediction problems across fields. For an individual member $i$ of a population $\mathcal{I}$ with covariates $X_i$, this can be expressed by the model $P_i = f(X_i)$ for $f : \mc X \to [0,1]$.  Standard statistical modeling approaches, such as logistic regression, are often used to estimate $f$ (see, e.g., \cite{hennepin, CPAT, lowenkamp2009development, vannostrand2009pretrial, levin2012santa}).
To facilitate easy manual computation of the scores, sometimes the regression coefficients are transformed or restricted to integer values \citep{jung2020simple, zeng2015interpretable}. 

Before being displayed to criminal justice decision-makers, the predicted probabilities, $\hat{P}$, under the estimated model are often translated to coarser risk groups, $g \in \mathcal{G}$, by binning the estimated probabilities (see, e.g., \cite{levin2012development, CPAT}). For example, the lowest risk group may be defined as including all individuals for whom $0 \leq \hat{P}_i < c_1$ for some policy-maker-determined value $c_1$. Similarly the $k$th group could be defined as being all individuals for whom $c_{k-1} \leq \hat{P}_i < c_k$, for some number of risk groups, $k=1,...,K$. Though equivalent, in practice the bins may be defined with respect to some monotone transformation of the $\hat P_i$s (e.g. the un-transformed linear predictor in a logistic regression with integer coefficients). We use the notation $\tilde{g}(\cdot ; c)$ to denote the function that bins predictions to risk groups. When the mapping is done using $\hat{P}_i$, we write $\tilde{g}( \hat{P}_i; c) = g$ to mean that individual $i$'s point estimate $\hat{P}_i$ places them into bin $g$ according to thresholds $c$.

\begin{figure}[ht]
\centering
\includegraphics[width=0.5\textwidth]{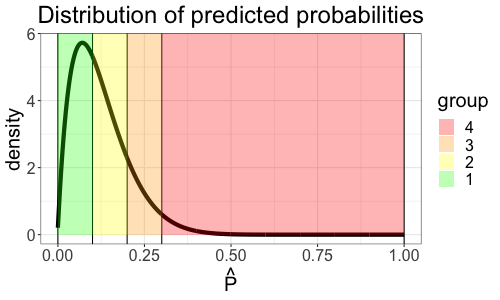}
\caption{\label{fig:example-groups} Hypothetical example of how individuals are partitioned into risk groups based on a model's predicted value or score.}
\end{figure}

A cartoon version is shown in Figure \ref{fig:example-groups}, in which the distribution of the predicted probabilities $\hat{P}$ from some hypothetical model are shown. Individuals are divided into risk groups indicated by different colored regions.  Risk groups may be numbered as shown (see, e.g., COMPAS decile scores discussed in the validation study of \cite{farabee2010compas}). They may also be given qualitative labels. For example, individuals whose predicted probabilities fall within the light red region could be labeled as ``high" risk. Those in the light orange, yellow, and green regions may be labeled ``medium-high", ``medium-low", and ``low" risk, respectively.

The risk groups may be used simply to inform the decision-maker about the estimated probability that an individual will have the outcome $Y$ or, in some cases, the risk groups are used to prescribe policy or treatment recommendations. For example, assignment to the highest risk group in one RAI used in New York City makes defendants ineligible for the supervised release program \citep{lum2019measures}.  Elsewhere, the PSA, which is commonly used for pre-trial decision making, often maps risk groups for three separate outcomes (failure to appear for court dates, re-arrest, and violent re-arrest) to recommendations for conditions of release using a Release Conditions Matrix \citep{psa-dmf}. Similarly, the Virginia Pretrial Risk Assessment Instrument uses the ``Praxis'' to map recidivism predictions to release recommendations. Both the Release Conditions Matrix and the Praxis are structured decision-making tools that combine the risk group assignment and charge information to recommend various levels and types of pre-trial supervision.\footnote{See the VPRAI instruction manual \href{https://www.dcjs.virginia.gov/sites/dcjs.virginia.gov/files/publications/corrections/virginia-pretrial-risk-assessment-instrument-vprai\_0.pdf}{here}} Notably, in Kentucky--- the state from which we obtained our data--- although the PSA is used, risk scores or groups are not associated with particular policy recommendations via the Release Conditions Matrix.
 
\section{Quantifying uncertainty in risk assessment models}
Given that the model's predictions and group classifications affect highly consequential decisions determining an individual's liberty, questions surrounding the uncertainty of the predictions and classifications naturally arise.  However, how one should formulate the problem of uncertainty quantification in this context--- uncertainty {\it about what}, so to speak --- has been a point of contentious debate 
(see, for example, \cite{hanson2017communicating, hart2007precision,imrey2015commentary, dawid2017individual}). Here, we discuss three types of uncertainty quantification relevant to risk assessment in increasing granularity: model-level, group-level, and individual-level. 

\subsection{Model-level}
At the {\bf model-level}, attention rests on evaluating the predictive utility of the model on the aggregate.  At this level, emphasis is on establishing that the model's predictions are positively correlated with the outcome of interest beyond what would be expected to occur due to chance alone. Interest also lies in establishing that the factors or covariates included in the model are statistically significant marginal correlates of the outcome. This is sensible from the point of view that one might reasonably demand that any factor that is used to justify a person's assignment to a particular risk group or score be based on scientific evidence that the factor is actually related to the outcome in question. Further, one might equally ask the question of whether the use of all of the factors in combination are actually, to a reasonable degree of statistical certainty, correlated with the outcome. 


This type of analysis is often performed via traditional statistical testing or interval estimation. For example, several manuals on the development of risk assessment tools report common predictive performance metrics, such as area under the receiver operating characteristics curve (AUC), with associated uncertainty intervals or measures of statistical significance for the model's regression coefficients (e.g. see \cite{vannostrand2003assessing, lowenkamp2009development, levin2012development, vannostrand2009pretrial}). Sometimes, bivariate tests of dependence between each covariate and the outcome variable are also reported (see, e.g. \cite{levin2012development}). These metrics speak to the model's overall predictive utility and whether each of the included factors are related to the outcome of interest. However, this sort of analysis does not preclude the possibility that the rate of failure is statistically indistinguishable between risk groups defined by the model's predictions.

\subsection{Group-level}
Uncertainty quantification at the {\bf group-level} pertains to uncertainty about sub-groups of individuals. Typically those sub-groups are defined by risk groups $\mathcal{G}$. 
Risk assessment tools can be viewed as a data-driven method for defining policies that partition individuals into groups, where the same treatment is recommended for all individuals within the same group. Recent policy simulations in \cite{kleinberg2017human} and \cite{picardbeyond}
have suggested that if risk assessment were used in place of human judgment to determine who would be released (i.e. all people within the same group received the same treatment), more individuals could be released while maintaining current rates of pre-trial failure. Perhaps from the point of view of a policy maker whose focus is on the impact of a proposed policy, the relevant question might then pertain to $P_g$--- the rate at which the outcome occurs in each group $g$. That is, if the tool were the instantiation of a policy for making release decisions, how uncertain would we be about the expected rate of the outcome among those who would be released under the policy? 

Indeed, most group-level uncertainty analysis in risk assessment has focused on uncertainty intervals for $P_g$. For example, \cite{glover2017cross} reports 95\% confidence intervals for the proportion of individuals who were re-arrested within each risk group  as defined by the tool, evidently calculated using a normal approximation to the binomial distribution. 

\begin{figure}
    \centering
    \includegraphics[width=0.5\textwidth]{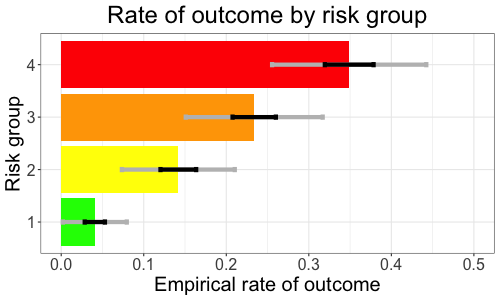}
    \caption{Example group-wise analysis. Bar lengths indicate point estimates of the FTA rate across all individuals classified to each group. Gray error bars show expected length of classical 95\% confidence intervals for data generated with 100 observations per group; black error bars show the equivalent for 1000 observations per group.}
    \label{fig:example-group-uq}
\end{figure}

\cite{hanson2017communicating} reports recidivism rates and confidence intervals for each risk group for the STATIC-99R, a risk assessment tool used to classify sex offenders into risk groups. \cite{demichele2018public} report boostrapped 95\% confidence intervals for the rate of FTA, new arrest, and new violent arrest for each of the risk groups in the PSA. \cite{scurich2012bayesian} purports to take a Bayesian approach to obtain individual-level credible intervals, though in doing so, implicitly assumes all individuals within the same risk group have the same $P_i$. Notably, these sorts of uncertainty intervals shrink as the sample size grows, as observing more individuals leads to more certainty about the {\it average outcome} within the groups. This is shown in Figure \ref{fig:example-group-uq}, which shows a prototypical group-wise analysis. Bar lengths show the observed rate of the outcome in the data for people classified to each group. Error bars show expected classical 95\% confidence intervals for data simulated with 100 individuals per group (gray) and 1000 individuals per group (black). Thus, while these intervals are informative about the average rate of the outcome within each group, they have more to do with the sample size used in the validation study than they do with defining a range of plausible values for each individual in the group. That is, these sorts of analyses say little about uncertainty regarding each individual's propensity toward the outcome, only the uncertainty about the average across individuals within each group. 

\subsection{Individual-level}
This framing raises the question of {\bf individual-level} variability within risk group and uncertainty quantification for individual-specific probabilities. \citet{hart2007precision} and \citet{hart2013another} have argued that, for risk assessments to be meaningfully applied to individual judicial decisions, uncertainty quantification at the individual level is of the utmost importance. While the statistical arguments employed by \citet{hart2007precision} to support this assertion were ill-formulated \citep{imrey2015commentary}, the normative argument about the importance of estimates of and uncertainty quantification for $P_i$ remains relevant.

\begin{figure}
    \centering
    \includegraphics[width=5in]{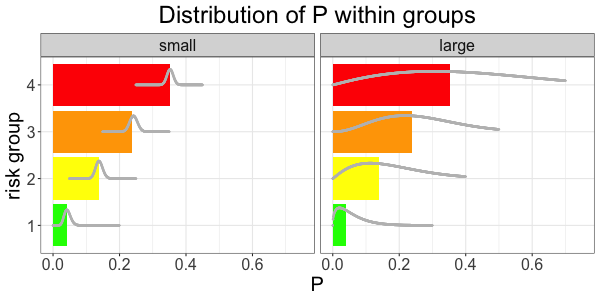}
    \caption{Two possible scenarios. The colored bars show the mean probability of the outcome by for each risk group, while the gray densities indicate the distribution of $P_i$ for individuals within each risk group. The left panel shows a case where everyone within each risk group has a very similar probability of the outcome; the right panel shows individual-level probabilities highly dispersed within each risk group.}
    \label{fig:indiv-dists}
\end{figure}

To formalize a bit, one might ask: given that an individual has been assigned to group $g$, how different might their individual probability $P_i$ be from the group probability $P_g$ attributed to them?  How certain are we about their $P_i$? Given the information available, what is the range of plausible values their individual $P_i$ could take? For $g(i)$ the group to which individual is $i$ is assigned, is the implicit assumption that $P_i = \hat{P}_{g(i)}$--- a simplification termed {\it individualized} risk \citep{dawid2017individual}--- approximately true or is it the case that some individuals within the same group have very different probabilities of the outcome than others?  These two different scenarios are shown in Figure \ref{fig:indiv-dists}. In the left panel, the distribution of $P_i$ is tight, and individuals within the same risk group all have essentially the same probability of the outcome. In the right panel, the distribution of $P_i$ is fairly diffuse around the group-wise mean. In this scenario, individuals in the same group can have very different individual level probabilities of the outcome. Their personal $P_i$ can be closer to the group-wise rate of the outcome for other risk groups than it is to the one to which they were assigned. 

In both of these scenarios, the marginal rate of the outcome and the observed data are exactly the same. However, in an ideal world in which knowledge of these true underlying distributions were available, these scenarios may support substantively different decisions. For example, suppose the small variance scenario in the left panel of Figure \ref{fig:indiv-dists} is true and a judge is presented with a case in which the person who has been assessed falls into risk group $g$ where $P_g = 0.35$ (the highest risk group in Figure \ref{fig:indiv-dists}). If the judge also knows that all individuals in group $g$ have individual probability $P_i \approx P_g = 0.35$ of experiencing the outcome (the case in the left panel), they might be willing to deny release for all individuals. They might believe that every person in this group is deserving of detention by virtue of the fact that they would fail to appear on 35\% of all similar occasions at which they were released. Suppose instead the large variance scenario in the right panel of Figure \ref{fig:indiv-dists} is true. Although the group-wise average remains $P_g = 0.35$, it is also known that about a quarter of the people in this group have individual probability less than $0.2$, and another quarter have probability greater than $0.5$. The individual probabilities are highly dispersed about the mean. The judge then might not be willing to detain individuals in this group on the basis of this risk score. The judge might believe that--- although they don't know which individuals are those with probability less than $0.20$--- individuals who would only fail at most 20\% of the time are not deserving of detention, even if releasing them also means releasing some individuals who would fail 50\% of the time. The judge might consider a 25\% chance of detaining someone whose personal probability of the outcome is only 20\% to be unacceptable.

 Unfortunately, the world is not ideal and we do not directly observe the underlying distribution of $P_i$s. Instead, we observe binary $Y_i$s. When only the $Y_i$s are observed, traditional group-wise analysis can't differentiate the two scenarios. Whether the underlying truth is the ``large" or ``small" variance scenario, group-wise rates and associated classical confidence intervals under both scenarios are exactly the same for the same sample size. 
 
  What, then, can be said about individual probabilities? On a purely mechanical level, it is possible to derive uncertainty intervals for $\hat{P}_i$ from model-level analysis. For example, consider a standard generalized linear model as in equation \eqref{eq:glm}.

\be \label{eq:glm}
    Y_i &\sim \Bern(P_i), \quad P_i = f(X_i\beta) 
\ee
Here, $f$ is a function mapping $X_i$ to $[0,1]$, such as the probit or inverse logit functions; we use the probit throughout this paper. Under this model, intervals for $P_i$ can be obtained as a function of the sampling distribution of $\hat \beta$.  Similar to standard estimates of uncertainty for $\hat{P}_g$,  intervals for $\hat{P}_i$ under equation \ref{eq:glm} shrink with increasing sample size, as the sampling distribution of $\hat \beta$ gets tighter with increasing $N$, the total number of observations. Such intervals are unsatisfying as measures of uncertainty at an individual level, as implicit in the model is the assumption that any two individuals with the same measured covariates necessarily have the same individual probability of the outcome. That is, $P_i$ as defined in this model does not correspond to a notion of individual probability that recognizes that individuals may differ along dimensions that are not captured by covariates. Any systematic variation across individuals that is not captured by any of the measured covariates is assumed away by the model's structure. Estimates of $P_i$ under this model, and uncertainty quantification thereof, then pertain to estimates of a quantity that by definition is not a meaningful mathematical embodiment of individual probability as one's personal, individual-specific probability of the outcome.

This raises the question of what even is an individual-specific probability of an outcome that only occurs once and how does one write down a model to estimate such a thing? This quickly gets philosophical. Our approach embraces a ``personalistic'' notion of probability described by \cite[pg. 3471]{dawid2017individual}, which is essentially Bayesian notion of probability. Fortunately, while the philosophical issues surrounding the notion of personalized probabilities are esoteric, contentious, and difficult, the \emph{implementation} of such an approach is, methodologically, very standard. Taking a personalized approach to the notion of individual probabilities essentially means doing Bayesian analysis. In practice, Bayesian analysis involves specifying a probabilistic model that incorporates all of the information at our disposal about the process that generates the data. This includes not only the choice of prior hyperparameters, but also distributional assumptions and choices about how to hierarchically structure the model in order to induce patterns of conditional independence that are  reasonable given the process being modeled. While frequentist approaches to analogous models are possible, the Bayesian framework we propose eases computation and interpretation of uncertainty. In the next section, we construct such a model for FTA that is a natural choice for the longitudinal data that we seek to analyze.  Importantly, our model incorporates parameters that can fairly be described as pertaining to an individual's personal probability in that they can vary in individual-specific ways not described by covariates.

\section{A Bayesian hierarchical random effects model}
For the reasons outlined in the introduction, standard regression approaches to criminal justice risk assessment are not structured to estimate $P_i$ that enable interpretation as pertaining to individuals. Hierarchical random effects models, however, are suited to addressing this problem by explicitly modeling individual-level heterogeneity. Hierarchical models are a central feature of applied Bayesian statistics, and are covered in detail in the well-known text \cite[Chapter 5]{gelman2013bayesian}.

Random effects models incorporate individual-specific parameters to allow for variability in expected outcomes across individuals that are not explained by covariates \citep{gelman2013bayesian}. For the parameters of the random effects distribution $f_\eta$ to be identifiable with binary outcome data, it is necessary to have repeated observations on at least some subjects $i$ in the data.
Using this framework, we expand the canonical model in \eqref{eq:glm} by adding individual-specific intercepts $\theta_i$ that are drawn from a common distribution $f_{\eta}$
\be \label{eq:logreg}
Y_{ij} &\sim \Bern(P_{ij}), \quad P_{ij} = f(X_{ij}\beta + \theta_i), \quad i=1,\ldots,m, \quad j = 1,\ldots,n_i \\
\theta_i &\stackrel{iid}{\sim} f_{\eta},
\ee

where $f_{\eta}$ is some distribution parameterized by $\eta$. Here, $j$ indexes an individual's observations, with individual $i$ having been observed on $n_i$ occasions. Then, $Y_{ij}$ for $j=1,\ldots,n_i$ is the outcome for the $i$th individual on the $j$th occasion they were observed. $X_{ij}$ are the covariates of person $i$ at the $j$th occasion on which they were observed. Under this model, $\theta_i$ represents individual $i$'s individualized propensity toward the outcome. Though their estimated probability may change across occasions as a function of covariates, $\theta_i$ represents their systematic deviation from what would be expected based on the measured covariates alone. If the distribution of $\theta$ has small variance, then this indicates that individuals' probabilities do not deviate much from their expected value given covariates. If $\theta$ has large variance, this suggests that there is substantial between-person variability beyond what is explained by the covariates in the model.

To complete a fully Bayesian model, we assign priors to the regression coefficients $\beta$ and the parameters $\eta$ of the random effect distribution
\be \label{eq:priors}
\beta &\sim N(\beta_0, \Sigma),\qquad 
\eta \sim \pi_{\eta}.
\ee
The prior distribution on $\beta$ reflects our prior beliefs about the relationship between the covariates in $X$ and the outcome. The $\beta$ parameters are common across all individuals in the data. We select prior hyperparameters $\beta_0 = 0$ and $\Sigma = 9 I$--- each $\beta$ parameter has an independent N(0,9) prior. We select prior variances of nine  because we standardize $X$ before computation, and an effect size of six is very large on the probit scale. We do not expect to see effects any larger than that.

By placing a prior on the random effect parameters $\eta$, we learn the distribution of $\theta_i$ across individuals in the population from the data. To ensure that our qualitative conclusions are robust to the distributional assumptions about the random effects, we consider two different forms for $f_{\eta}$. The first is a single normal distribution, $\theta_i \sim N(\mu, \tau)$. The second is a discrete mixture $f_{\eta}(\theta_i) = \sum_{k=1}^K w_k \delta(\theta_i - \mu_k)$, where $\mu_k \sim N(0,1)$ and $(w_1, ..., w_K) \sim \text{Dirichlet}(1/K, ..., 1/K)$. 
We emphasize the results for the single normal distribution in the main text but supply full analysis of the discrete mixture in the appendix. The results vary little between the two models.

\section{Computation}
For computational convenenience, we use the probit link to specify $f$, and express the model in the uncentered parametrization. When the data are only weakly informative about the $\theta_i$, the uncentered parametrization leads to more rapid convergence of Markov chain Monte Carlo (MCMC) algorithms of the type that we employ in this paper \citep{papaspiliopoulos2007general}. We consider the case where $f_\eta$ is the density of a normal distribution and the parameter $\eta = (\mu,\tau)$ is a vector of the mean and scale of the random effects distribution.

Our complete model in the uncentered parametrization is
\be
y_{ij} &\sim \Bern (\Phi(\mu + \tau \theta_i + x_{ij} \beta) ) \\
\theta_i &\stackrel{iid}{\sim}  N(0,1) \\
\mu &\sim N(0,9), \quad \tau \sim N_1(0,1), \quad \beta &\sim N(0,9 I),
\ee
where $N_1$ denotes a normal distribution truncated to the positive half-line and our prior for the standard deviation $\tau$ of the random effects distribution is weakly informative. We use a data augmentation blocked Gibbs sampler for computation. In our data, the total number of observations is large ($N := \sum_{i=1}^m n_i$ is in the hundreds of thousands) while the number of individuals $m$ is also in the hundreds of thousands, since most individuals have $n_i = 1$ or $n_i = 2$. In this setting--- many observations but few observations per individual--- we have found general-purpose probabilistic programming tools such as \texttt{Stan} to have very high computational cost per iteration, and that blocked data augmentation Gibbs sampling in the uncentered parametrization offers by far the best performance of commonly used algorithms when accounting for both computational cost per step and convergence rate of the MCMC algorithm. Furthermore, this approach can be used with some modifications for both the Gaussian and discrete mixture models for $f_\eta$. The updates for the Gaussian specification of $f_\eta$ are as follows: 
\begin{enumerate}
\item Update $\omega_{ij}$, all conditionally independent, from
\be
\omega_{ij} &\sim N_{y_{ij}}(\mu + \tau \theta_i + x_{ij} \beta, 1),
\ee
where $N_y$ is a normal distribution truncated to $(0,\infty)$ if $y$ is positive and is a normal distribution truncated to $(-\infty,0]$ if $y$ is nonpositive.

\item Update $\tau$ marginal of $\mu$ from
\be
\tau \sim N_1\left( \alpha_\tau, s_\tau \right)
\ee
where
\be \label{eq:TauFC}
s_\tau &= \left( \sum_{i=1}^m (n_i \theta_i^2) - \left(\sum_{i=1}^m n_i \theta_i \right)^2 (1/9 + N)^{-1} + 1  \right)^{-1} \\
\alpha_\tau &= s_\tau \left(\sum_{j=1}^{n_i} \sum_{i=1}^m \theta_i(\omega_{ij}-x_{ij}\beta) - \left( \sum_{i=1}^m n_i \theta_i \right) \left( \sum_{i=1}^m \sum_{j = 1}^{n_i} \omega_{ij}-x_{ij} \beta \right) (1/9+N)^{-1} \right).
\ee
Derivation of this full conditional is given in the Supplementary Materials. 
\item Update $\mu$ given everything from 
\be
\mu &\sim N(\alpha_\mu,s_\mu) \\
s_\mu &= (1/9 + N)^{-1}, \quad \alpha_\mu = s_\mu \left( \sum_{i,j} \omega_{ij} - \tau \theta_i - x_{ij} \beta \right)
\ee

\item Update $\theta$ given everything from 
\be
\theta_i &\sim N(\alpha_i,s_i) \\
s_i &= (\tau^2 n_i + 1)^{-1}, \quad \alpha_i = s_i \sum_{j=1}^{n_i} \tau (\omega_{ij} - \mu - x_{ij} \beta)
\ee

\item Update $\beta$ given everything from
\be
\beta &\sim N(\alpha_\beta,S_\beta) \\
S_\beta &= (X'X + (1/9) I)^{-1},\quad \alpha_\beta = S_\beta X'(\omega - \mu 1_N - \tau W \theta).
\ee
\end{enumerate}
We run the algorithm for 20,000 iterations, discarding the first 5,000 iterations as burn-in. Computation details for the discrete mixture distribution are  given in section \ref{sec:disc-mix-comp} of the Supplementary Materials.

\section{Data}
We obtained data from Kentucky pre-trial services covering all people who were arrested and booked in the state from July 1, 2009 through the end of 2018. This consists of nearly 1.5 million arrests. Associated with each arrest is an ``interview" at which time information about the person who was arrested was collected, including their age, criminal history, and history of failing to appear for court appointments. One interview can be associated with multiple cases. Here, the unit of analysis is an arrest (or, equivalently, interview). In June of 2014, the information recorded at each interview changed when a new risk assessment model was introduced. In order to build a coherent regression model using a consistent set of covariates, we restrict the dataset to only include interviews that took place after June 30, 2014. This leaves approximately $\numprint{611000}$ arrests for our analysis. 
We restrict the data to only those interviews for which the individual was at any time released pre-trial, since there is no opportunity for FTA if the individual is never released. This encompasses around $\numprint{463000}$ arrests and associated pre-trial outcomes. We drop $\numprint{72}$ observations due to missingness. We reserve all arrests that took place in 2018, the last year of our data, as a holdout set on which to test our model. The holdout set makes up $\numprint{13}$\% of our data.

\begin{figure}[ht]
\centering
\includegraphics[width=3in]{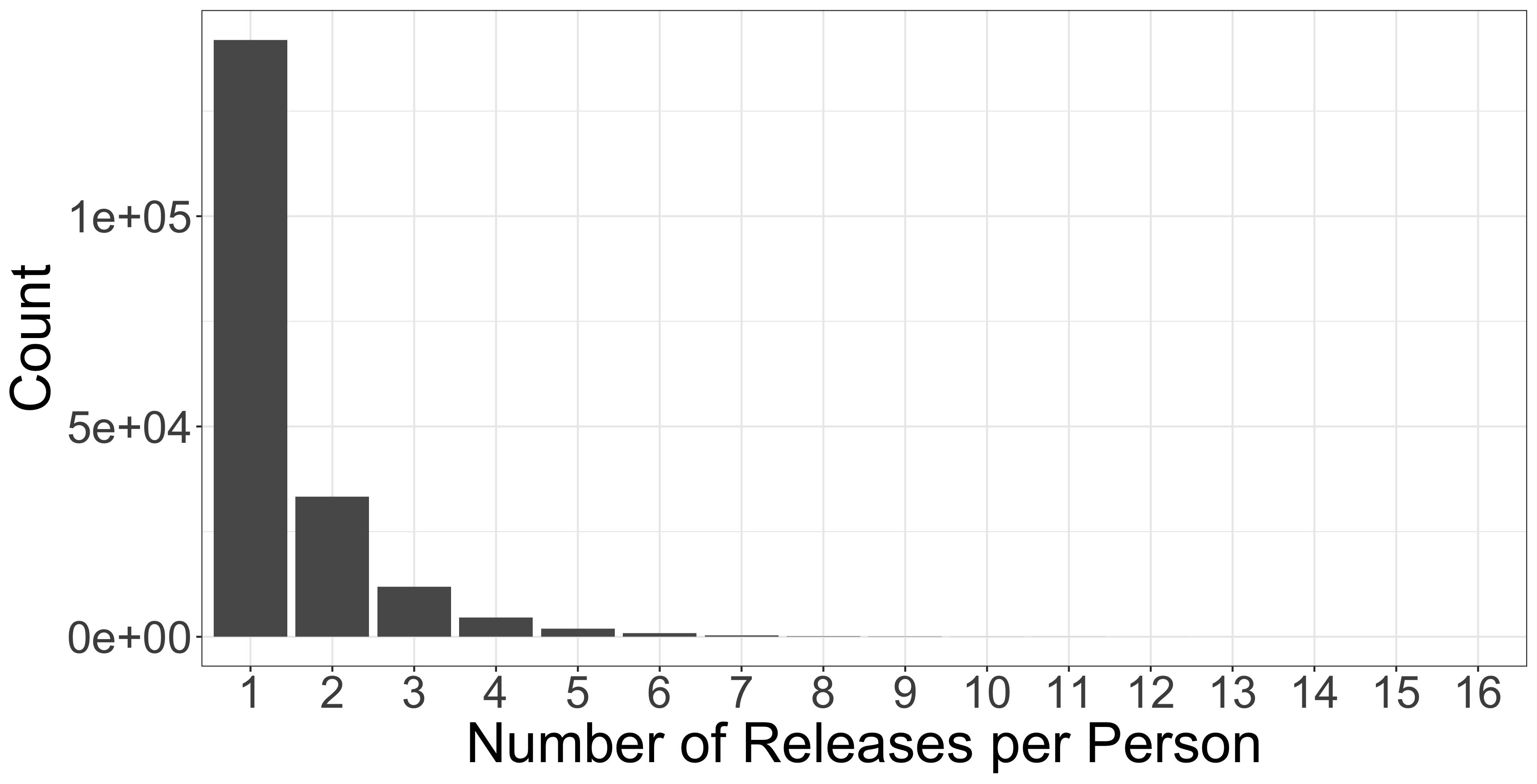}
\caption{\label{fig:disagg2} The distribution of the number of times each individual appears in the training dataset.}
\end{figure}

Figure \ref{fig:disagg2} shows the distribution of number of releases per-person in the training dataset. The plurality of individuals (around $\numprint{142000}$) had only one arrest with release during the time period covered by our training data.

Associated with each arrest are several measures recorded at the time of the pre-trial interview. The factors we use in our model are: whether the current charge is a domestic violence charge, whether the current charge is for a violent offense, whether the current charge is for a violent offense and the arrested person is under 21 years of age, the arrested person's age at the time of the interview, whether there is currently another case pending, whether the individual has had any prior misdemeanor arrests, whether they have had any prior felony arrests, the number of failures to appear for court appointments within the last two years, whether they had any failures to appear prior to two years ago, the number of prior arrests with violent charges, whether the individual has a prior incarceration, and whether at the time of the current arrest there was an outstanding warrant for an FTA. The measure of past failures to appear encompasses failures to appear for non-arrest-related court appointments, such as traffic violations, as well as failures to appear that occurred in other jurisdictions. This justifies its inclusion as a covariate, since the outcome variable in our model only records arrest-related FTAs. 

These factors make up the columns of $X$ and are a super-set of the factors used in the FTA risk model currently used to evaluate people who have been arrested and inform pre-trial decisions in Kentucky. Associated with each record is also a FTA risk score that was generated at the time of the interview. These risk scores take values one through six, where larger values indicate a higher estimated likelihood of failure to appear. Each numeric score is considered a different risk group. As of November 2017, Kentucky uses a point scale that runs from zero to seven instead of the six groups considered here. Because the six point scale is commonly used and was used for the majority of the time period covered by our data, we use the 6 point scale throughout this analysis. We do not use these risk groups as covariates in our model, but we do use them for later evaluation.

The outcome variable used in our analysis is a binary indicator of whether the individual failed to appear for any court appointments for any case associated with their interview/arrest prior to their case disposition.\footnote{It may seem redundant to include this measure of FTA as the outcome and also include past FTAs as a covariate in this longitudinal setting. There are significant differences between the two measures: the outcome measure of FTA only pertains to FTAs for cases in this jurisdiction associated with arrests. The measure of FTA used as a covariate (which is calculated at the time of the interview) may encompass FTAs from other jurisdictions, including FTAs for violations. However, we fit the model excluding past FTAs as a covariate and had qualitatively similar findings. The results without FTA as a covariate paint a much less optimistic picture of the potential for risk assessment to create meaningful individual-level differences, as the signal in the covariates is substantially decreased.} This outcome variable has some conceptual limitations. Though there may have been several required appointments, a single missed appointment results in the failure to appear indicator. This measure of pre-trial failure does not disambiguate between individuals who nearly always appear as required and those who have absconded or habitually miss appointments. However, this definition of the FTA outcome variable is frequently used for failure to appear prediction and model evaluation (see, e.g., \cite{demichele2018public}, for a recent example), and the data required to calculate this alternative measure of FTA (all of the court dates to which the individual did appear as scheduled) is not available. The rate of FTA in these data is $20.7$\%.

\section{Results}
Here we present results for each of the ``levels" of analysis in turn. Results under the normal and discrete random effects distributions gave qualitatively similar results, with the discrete distribution exhibiting a slightly longer right tail for the distribution of individual probabilities. Because the results were qualitatively similar, we focus here only on the results for the normally distributed random effects. The lower-dimensional parameterization of this model allows for ease of exposition as the $\tau$ parameter offers a succinct and convenient way to discuss individual-level variability. Results for the discrete mixture distribution are given in appendix \ref{sec:disc-mix-results}.

\subsection{Model-level}
We begin by summarizing posterior inference for the global model parameters, $\beta$ and $\tau$. Table \ref{tab:beta} gives the posterior mean and 95\% posterior credible intervals for each of the regression coefficients used in our model. The posterior intervals for most coefficients do not include zero, though most coefficients have small effect sizes. 
Due to the sample size available for fitting this model, the posterior credible intervals are all quite tight around the posterior mean. 

\begin{table}
\caption{Posterior summary statistics of regression coefficients.} 
\label{tab:beta}
\begin{tabular}{llll}
  \hline
 & mean & 95\% CI & description \\ 
  \hline
DV Charge & -0.07 & (-0.08,-0.04) & Current domestic violence charge \\ 
  Current Violent & -0.27 & (-0.28,-0.25) & Current violent charge \\ 
  Violent x Age & -0.08 & (-0.09,-0.03) & Current violent charge and person under 21 years old \\ 
  Age 21 or 22 & 0.07 & (0.07,0.1) & Aged 21-22 at time of interview (baseline under 21) \\ 
  Age 23+ & 0.11 & (0.11,0.14) & Aged 23+ at time of interview (baseline under 21) \\ 
  Pending Cases & 0.19 & (0.18,0.2) & Any pending cases \\ 
  Misdemeanors & 0.13 & (0.12,0.14) & Any prior misdemeanor convictions \\ 
  Felonies & 0.07 & (0.07,0.09) & Any prior felony convictions \\ 
  FTA 1-2 & 0.18 & (0.17,0.19) & 1 or 2 FTAs in the last two years (baseline 0) \\ 
  FTA 3+ & 0.28 & (0.28,0.3) & 3+ FTAs in last two years (baseline 0) \\ 
  Old FTAs & 0.14 & (0.13,0.15) & Any FTAs older than two years \\ 
  Violent 1-2 & 0 & (-0.01,0.01) & 1 or 2 prior violent convictions (baseline 0) \\ 
  Violent 3+ & -0.02 & (-0.03,0.01) & 3+ prior violent convictions (baseline 0) \\ 
  Incarcerations & 0.12 & (0.12,0.13) & Any prior incarcerations \\ 
  Current FTA Case & 0.42 & (0.41,0.43) & Current FTA warrant associated with arrest \\ 
   \hline
\end{tabular}
\end{table}

\begin{figure}
    \centering
    \includegraphics[height=1.75in]{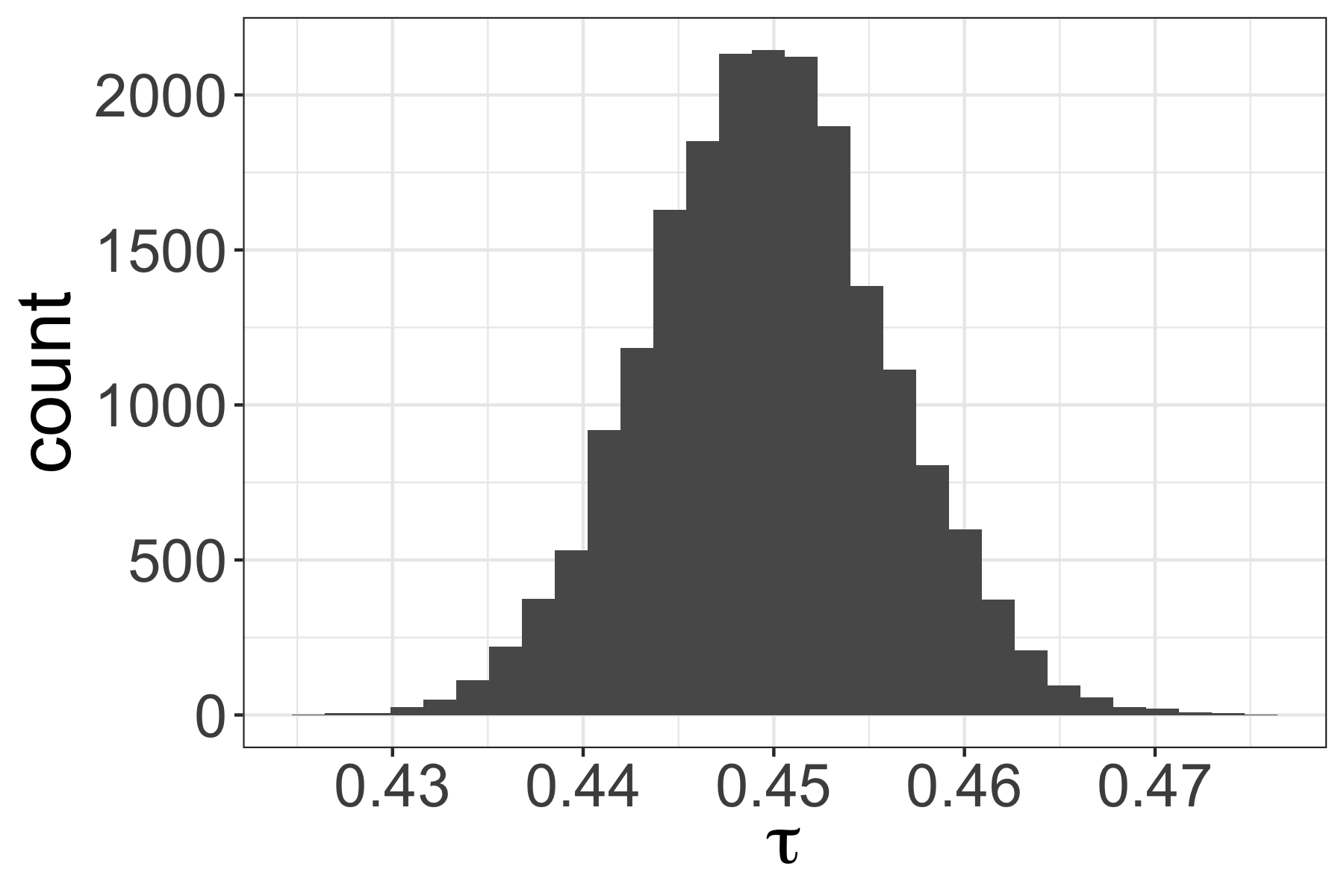}
    \includegraphics[height=1.75in]{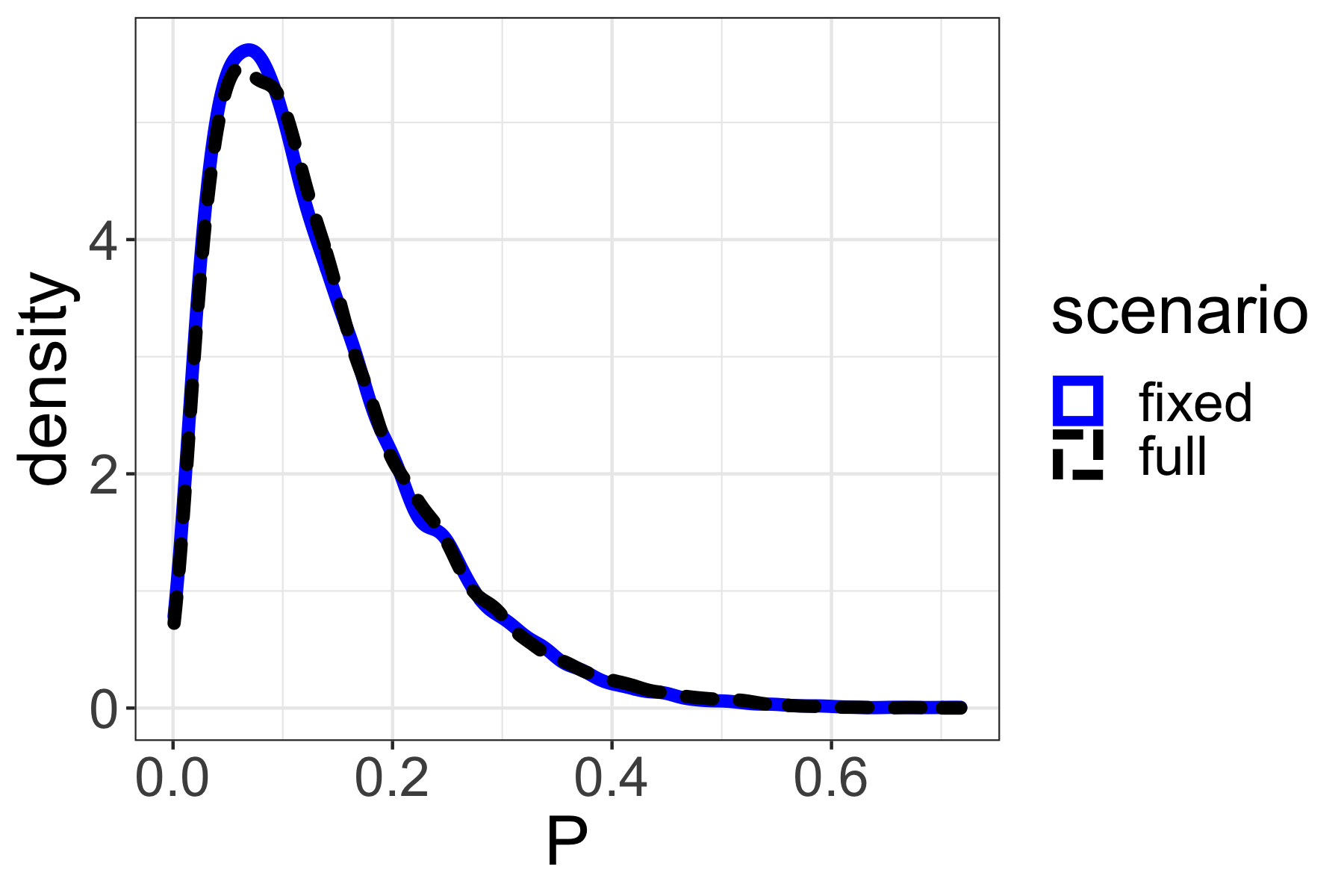}
    \caption{(left) Posterior samples of $\tau$ in the normal random effects model; (right) Density plots of samples of $P_{\text{med}}$ under the posterior predictive distribution (``full") and the distribution assuming fixed global parameters (``fixed").}
    \label{fig:tau-hist}
\end{figure}

The left panel of figure \ref{fig:tau-hist} shows a histogram of posterior samples of $\tau$ from the normal random effects model. This posterior distribution is fairly tight around its mean. By the law of total variance, posterior uncertainty about $P$ can be decomposed into $\var(P) = E(\var(P \mid\beta, \tau)) + \var(E(P \mid \beta, \tau))$, where the first term is the expected amount of `natural' variability in $P$ across the population after conditioning on model parameters, and the second term reflects uncertainty about the model parameters. We belabor this point because our posterior uncertainty about $P \mid X$ can be large for two reasons. We can have large uncertainty about $P$ because individuals' probabilities vary considerably about their expectation given covariates, i.e. $\tau$ is large. This scenario is morally equivalent to a model with large residual variance. One could also have large uncertainty for $P$ because the available data are insufficient to estimate the model parameters precisely. The latter case is in some ways less interesting, as learning that we simply have not collected enough data to estimate model parameters with any precision presents a less compelling case than learning that individuals are much more variable than we've previously recognized. The latter case could be remedied by collecting more of the same data, while remedying the former would require discovering heretofore unknown covariates that explain a significant portion of the unexplained variability. Or, it may be the case that $\tau$ is truly just large in the sense that after conditioning on all information that is legal or feasible to collect, people naturally vary widely in their propensity and ability to appear.

To decompose the posterior uncertainty of $P$ into the effects of between-individual variability and posterior uncertainty for global parameters, consider the hypothetical ``median individual" whose covariate values each marginally take the median value, $X_{\text{med}}$. We assess posterior predictive uncertainty for this hypothetical individual on the first time they are observed using two methods. In the first (called ``full"), we sample from the full posterior predictive distribution. That is, we sample from $p_{\text{full}}(P_{\text{med}} \mid \text{data}) = \int p(P_{\text{med}} \mid X_{\text{med}}, \beta, \tau) p(\beta, \tau \mid \text{data}) d\beta d\tau$. This distribution incorporates our posterior uncertainty about model parameters.  In the second (called ``fixed"), we ignore posterior uncertainty in the model parameters and  sample from $p_{\text{fixed}}(P_{\text{med}} \mid \text{data}) \sim p(P_{\text{med}} \mid X_{\text{med}}, \hat{\beta}, \hat{\tau})$, fixing model parameters at their posterior mean. A comparison of these distributions is shown in the right panel of figure \ref{fig:tau-hist}. In this figure, we see that there is virtually no difference in the posterior predictive distribution of $P_{\text{med}}$ in these two cases. 
Thus, the uncertainty in estimating model parameters has very little impact on our uncertainty in $P$.

\subsection{Group-level}
In our group-level analysis, we assess the variability of $P_{ij}$ within risk groups in our holdout set. In order to create risk groups, we consider the predictive quantity $P_{ij}^*$, which we define in \eqref{eq:pstar}. To match as closely as possible how our model would be used in criminal justice risk assessment, we condition the global parameters $\beta, \tau, \mu$ only on the training data, but the individual-level parameters $\theta_i$ are updated every time new data arrive for individual $i$. Specifically, for any $\zeta \in [0,1]$, let $A_{ij}(\zeta) = \{(\theta_i,\tau,\mu,\beta) : \Phi(\mu+\tau \theta_i + X_{ij} \beta) \le \zeta\}$, and let $y_{\train},X_{\train}$ be the training data (all data through December 31, 2017). The CDFs $H^*_{ij}(\zeta) \equiv \bb P[P^*_{ij} \le \zeta]$ and $H_{ij}(\zeta) \equiv \bb P[P_{ij} \le \zeta]$ are given by
\be
H^*_{ij}(\zeta) &= \int_{A_{ij}(\zeta)} p(\theta_i \mid y_{i,1:(j-1)}, X_{i,1:j}, \tau, \mu, \beta) p(\tau, \mu, \beta \mid X_{\train}, y_{\train}) d\theta_i\, d\tau\, d\mu\, d\beta \label{eq:pstar}\\
H_{ij}(\zeta) &= \int_{A_{ij}(\zeta)} p(\theta_i \mid y_{i,1:j}, X_{i,1:j}, \tau, \mu, \beta) p(\tau, \mu, \beta \mid X_{\train}, y_{\train}) d\theta_i\, d\tau\, d\mu\, d\beta.
\ee
In other words, when computing the distribution of either $P^*_{ij}$ or $P_{ij}$, we condition the global parameters on the training data, but update the conditional posterior of $\theta_i$ as more information becomes available in the test data. The quantity $P^*_{ij}$ does not condition on having yet observed the outcome for individual $i$ on occasion $j$. Thus, $P^*_{ij}$ is the predictive quantity that would be used in risk assessment for individual $i$ on occasion $j$. On the other hand, $P_{ij}$ does condition on $y_{i,j}$, and is essentially the posterior distribution of the probability of FTA for individual $i$ after observing occasions $1$ through $j$. It differs from the exact posterior in that the global parameters are conditioned only on the data up through the end of 2017. Because the conditional posterior $p(\tau, \mu, \beta \mid X_{\train}, y_{\train})$ is already quite concentrated (see, e.g., the right panel of figure \ref{fig:tau-hist}), these quantities are a close approximation to what would result if we updated the full posterior distribution of all model parameters every time a new observation is obtained. We are unaware of any criminal justice risk assessment instruments that are trained in an online fashion, and thus these ``partial information'' posterior predictive and posterior quantities reflect our best approximation of how a model like ours would be used in practice.

There are several ways in which risk groups could be defined. We consider three. The first is to use the PSA's FTA risk groups that were calculated at the time of the interview. These are the groups that were used to make pre-trial recommendations. These risk groups do not perfectly correspond to any set of thresholds, $c$, applied to our model's output, as these risk scores were produced by binning different probabilities than those obtained by our model. However, for evaluation, we will need to define approximate thresholds, $c_{\text{PSA}}$. The values of $c_{\text{PSA}}$ are set to be the mid-point between the empirical rates observed in each of the PSA's risk groups. 

We could also create our own groups using our model. Let $\hat P^*_{ij} = \bb E[P^*_{ij}]$ be the mean of $P^*_{ij}$. We bin the $\hat P^*_{ij}$s to create our own risk groups. We determine the bin thresholds using only the training data.  For consistency with the PSA's typical scoring, we create six bins. We consider one binning scheme in which we create bins so that the number of people in the $k$th bin is the same as the number of people in the $k$th PSA risk group. We call this ``PSA-sized groups.'' We also create groups by applying k-means clustering with $K=6$ clusters. This allows the data rather than the precedent set by the PSA to determine the size of each of the bins. We refer to the thresholds used to create each of these groupings as $c_{\text{PSA-sized}}$ and $c_{\text{clustered}}$. In summary, we consider three groupings. The first, $g_{\text{PSA}}(i, j)$, returns the value of the $i$th individual's FTA risk score as given in the data on the $j$th occasion on which they were observed. The second, $g_{\text{PSA-sized}}(i,j) = \tilde{g}(\hat{P}^*_{ij} ; c_{\text{PSA-sized}})$ returns the $i$th individual's risk group applying thresholds $c_{\text{PSA-sized}}$ to each individual's $\hat{P}^*_{ij}$ under our model.  Finally, $g_{\text{clustered}}(i,j) =  \tilde{g}(\hat{P}^*_{ij} ; c_{\text{clustered}})$ returns the $i$th person's risk group from applying thresholds $c_{\text{clustered}}$. 

Table \ref{tab:model-fit} shows the average $\hat P^*_{ij}$ and $\hat P_{ij}$ for each risk group in our holdout set. For comparison, we also show the empirical rate of the outcome within each group. We construct frequentist 95\% confidence intervals for the proportion in each risk group using the normal approximation to the binomial distribution (shown in parentheses by the group-wise rates). Applying this standard group-wise analysis to the three sets of risk groups shows that the risk groups exhibit different average rates of FTA with only one exception. Notably, for the PSA Risk Groups, there is virtually no difference in the empirical rate of FTA between the two highest risk groups, five and six. This is also reflected by the fact that the confidence intervals for these groups overlap. 

We also see that the model is generally well-calibrated--- estimates of $P^*$  and $P$ are very similar to the true rate of the outcome within each group. We find the calibration to be quite good notwithstanding that in late 2017 (the end of the period from which we draw our training data), Kentucky created a specialized unit to conduct the risk assessment that resulted in more thorough assessments of past pre-trial failure than had been conducted previously. Despite difference this caused in the way the that the training and test data was recorded, our model remained satisfactorily calibrated. 

\begin{table}
\caption{The average posterior mean of $\hat P^*_{ij}$, $\hat P_{ij}$, and the ``true'', empirical mean of $Y_{ij}$ within risk groups under three different grouping strategies.} \label{tab:model-fit}

\centering
\begin{tabular}{l|lll|lll|lll}
\toprule
    & \multicolumn{3}{c}{PSA Risk Groups} \vline & \multicolumn{3}{c}{PSA-Sized Groups} \vline & \multicolumn{3}{c}{Clustered Groups} \\
\midrule
 & P* & P & rate & P* & P & rate & P* & P & rate \\ 
  \hline
1 & 0.10 & 0.10 & 0.08 (0.08,0.09) & 0.08 & 0.08 & 0.07 (0.06,0.08) & 0.09 & 0.09 & 0.07 (0.07,0.08) \\ 
  2 & 0.13 & 0.13 & 0.12 (0.12,0.13) & 0.12 & 0.12 & 0.11 (0.11,0.12) & 0.12 & 0.12 & 0.11 (0.1,0.11) \\ 
  3 & 0.18 & 0.18 & 0.18 (0.17,0.18) & 0.17 & 0.17 & 0.17 (0.16,0.18) & 0.15 & 0.15 & 0.14 (0.14,0.15) \\ 
  4 & 0.26 & 0.26 & 0.26 (0.25,0.27) & 0.24 & 0.24 & 0.25 (0.24,0.26) & 0.19 & 0.19 & 0.2 (0.19,0.21) \\ 
  5 & 0.36 & 0.36 & 0.36 (0.35,0.36) & 0.37 & 0.37 & 0.35 (0.35,0.36) & 0.26 & 0.26 & 0.27 (0.26,0.28) \\ 
  6 & 0.41 & 0.40 & 0.36 (0.35,0.38) & 0.54 & 0.54 & 0.5 (0.48,0.52) & 0.42 & 0.42 & 0.4 (0.39,0.41) \\ 
   \hline

\end{tabular}
\end{table}

\begin{figure}
    \centering
    \includegraphics[width=5in]{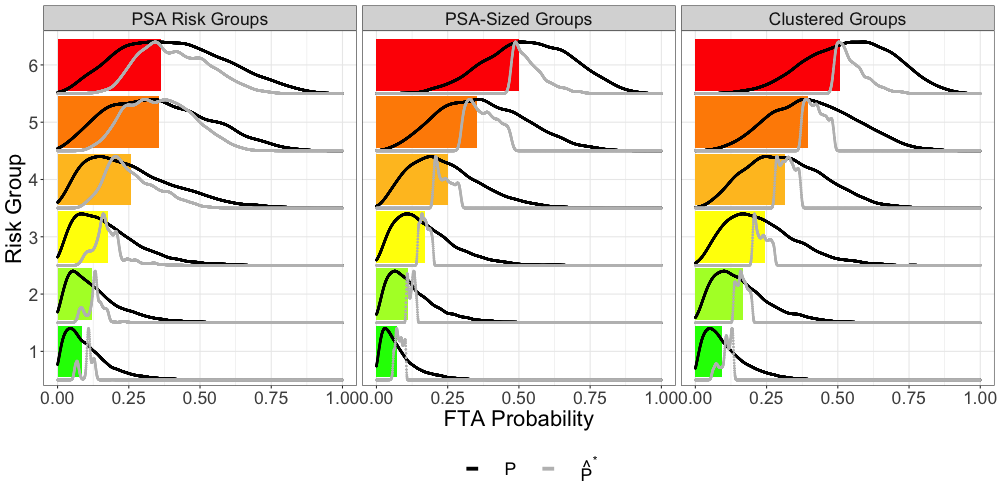}
    \caption{Posterior density of $P_{ij}$ and $\hat P^*_{ij}$ for each risk group (subscripts omitted in legend).}
    \label{fig:normal-psa}
\end{figure}

Figure \ref{fig:normal-psa} shows density estimates in black of the posterior distribution of $P_{ij}$ for each group, $g$, and grouping scheme, $s$: $p(P_{ij} \mid \text{data}, g_s(i,j) = g)$. The gray line shows a density estimate of $p(\hat{P}^*_{ij} \mid \text{data}, g_s(i,j) = g)$. In the former case, the density is the average over each observation in $g$'s posterior distribution of $P_{ij}$. In the latter case, each observation in group $g$ contributes one value, the point estimate of $P^*_{ij}$, to the density. For consistency with previous examples and displays, the empirical rate of FTA is shown by each of the colored bars. We find that the unexplained variation in the population within groups is substantial. Reality more closely resembles the``large" variance scenario than the ``small" variance scenario given in figure \ref{fig:indiv-dists}. The group-wise posterior distribution of $P_{ij}$ is broad, placing non-negligible mass in regions far from the group-wise mean in all cases. Returning to the discussion of the model-level analysis, the large posterior intervals within risk groups are attributable to having estimated with high certainty that there exists large unexplained between-individual variability, even after conditioning on available covariates.

\begin{figure}
    \centering
    \includegraphics[width=5in]{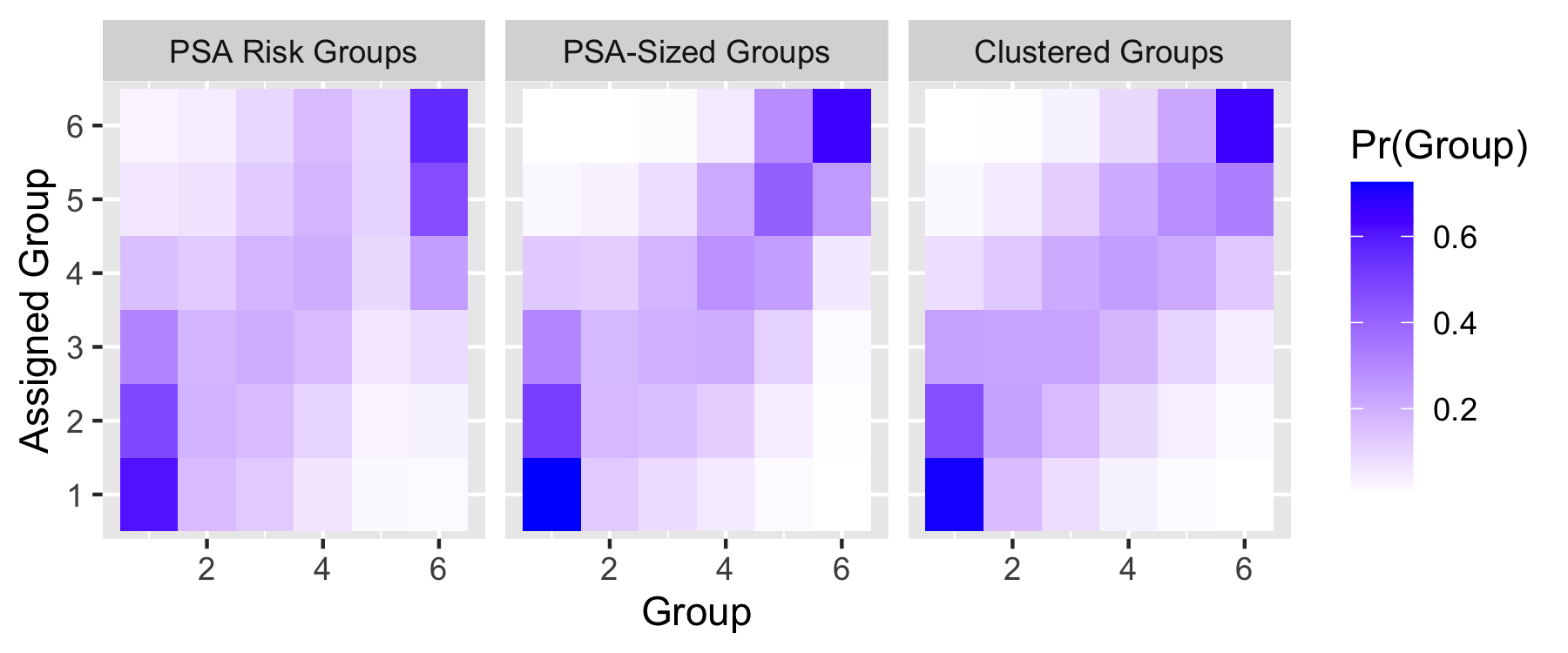}
    \caption{Posterior probability that $P_{ij}$ falls within the region of group $g$ for each actual group assignment ``Assigned Group".}
    \label{fig:wrong-bin}
\end{figure}

This motivates the question of how often an individual is assigned to the ``right" or ``wrong" risk group. To address this, we look at the distribution of $\tilde{g}(P_{ij}; c_s) \mid g_s(i,j)$ for each $s$--- the probability that a $P_{ij}$ falls into each bin, $g'$, conditional on having been assigned to group $g$ on the basis of their predictive point estimate. This is shown in figure \ref{fig:wrong-bin}. Each row indicates the assigned risk group for each of the three grouping schemes. The columns indicate each of the groups to which an individual could be assigned. The color in the $k$th row and the $k'$th column represents the posterior probability that a $P_{ij}$ falls within bin $k$ given that the  predictive point estimate assigns them to group $k'$. The first and last groups (columns) have the highest posterior probability because the range of values of $\hat P^*_{ij}$ corresponding to each of those bins is large relative to the middle groups. For example, while group one maps to values of $\hat P^*_{ij}$ in approximately the range $[0, .11)$, group two maps to values in the much smaller range of approximately $[.11, .15)$.\footnote{The precise thresholds vary depending on grouping scheme; these serve only as an example.} 

From this we see that assignment to group one is the most meaningful. For example, individuals assigned to group one under the PSA's risk groups have probability 0.61 of their $P_{ij}$ being within the range attributed to group one and probability 0.09 of it being in the range ascribed to groups four, five, or six. Similarly though less pronounced, people assigned to group six also had probability 0.56 of their $P_{ij}$ being in the range attributed to group six and probability 0.17 of $P_{ij}$ being in range associated with groups one, two, or three. Using group assignments derived from a model fitted to these data resulted in slightly better separation of individual-level probabilities into groups, though across all cases, there was substantial posterior probability of falling into some other group than the one assigned.  Specifically, the probability of $P_{ij}$ falling into the range attributed to the group to which it was assigned based on its predictive point estimate was 0.25, 0.32, and  0.39 for each of the binning strategies, respectively. This is because the proportion of arrests resulting in a group one or six assignment (those about which we are most certain) is fairly low. On most occasions, an individual receives a group two through five assignment.  The probability of such assignments having been correct tends to be low.  

\subsection{Individual-level}
\begin{figure}
    \centering
    \includegraphics[height=1.55in]{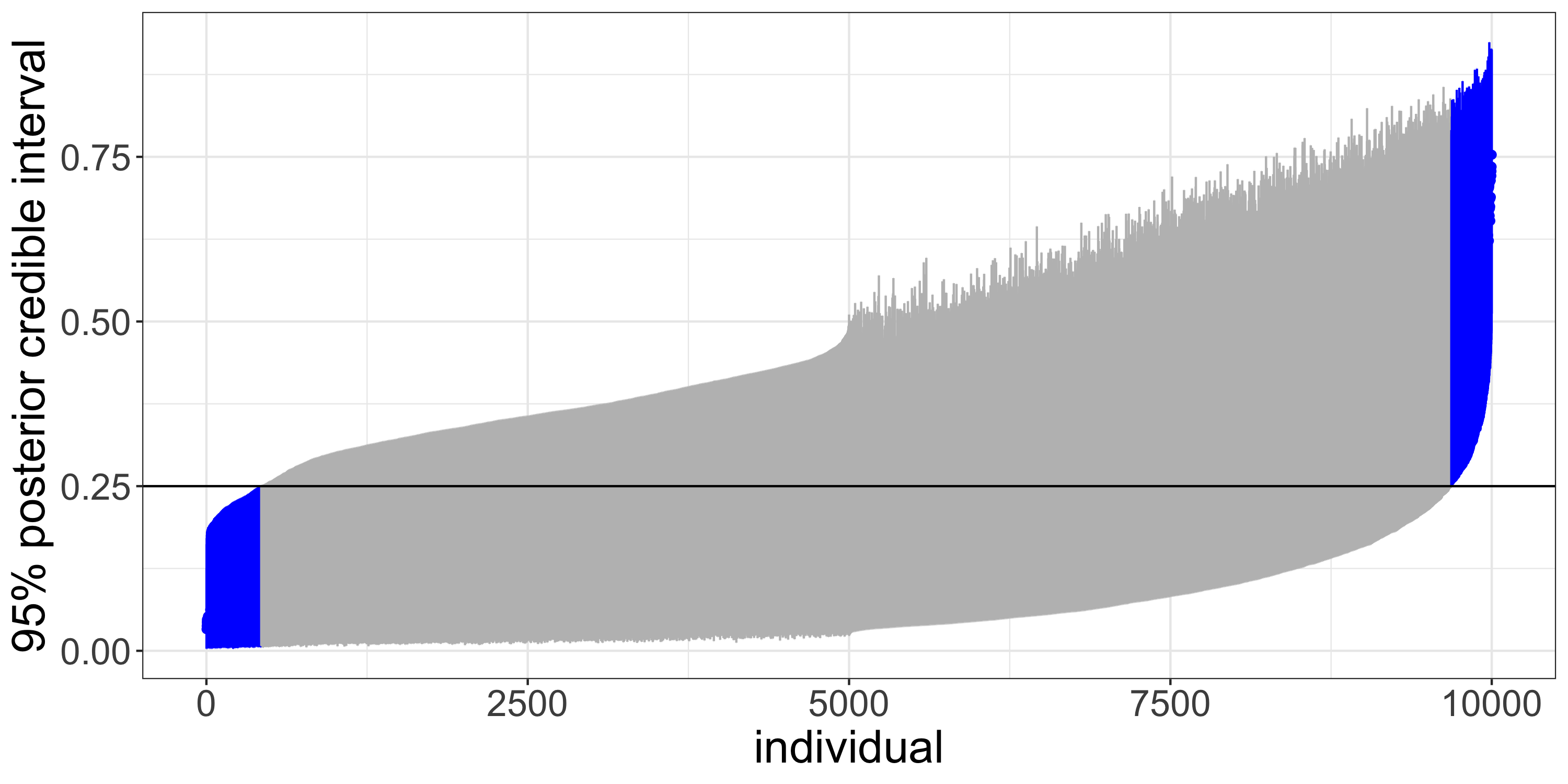}
    \includegraphics[height=1.55in]{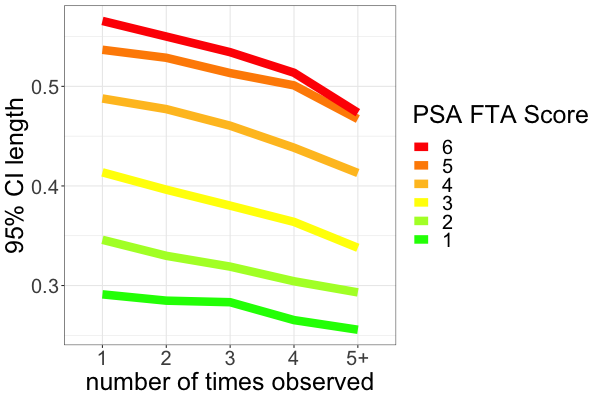}
    \caption{(left) 10,000 randomly selected 95\% posterior credible intervals for $P_{ij}$. Intervals shown in blue do not cross the 0.25 threshold. (right) Average length of 95\% posterior credible interval by number of times observed and PSA risk group.}
    \label{fig:interval-examples}
\end{figure}

Finally, we turn to explicitly evaluating individual probabilities and associated uncertainty by considering metrics based on individual posterior credible intervals for $P_{ij}$. The left panel of figure \ref{fig:interval-examples} shows a random sample of 10,000 95\% posterior credible intervals of $P_{ij}$. For each interval, we calculate the posterior median. For those intervals whose posterior median is in the bottom half of all intervals' medians, intervals are sorted by the upper end of the interval. For those whose posterior median falls in the upper half, intervals are sorted by the lower end of the interval.  This allows us to  approximate the extent to which individuals can be divided into groups that are statistically distinct from one another by determining whether their posterior credible intervals overlap. We choose a threshold of 0.25 and color intervals in blue if they do not cross that threshold. This creates two groups for which all of the individuals at the lower end of the distribution have intervals that do not overlap with those at the higher end. Under this strategy, one could flag 7\% of the arrests as belonging to one of the groups on the two ends of the spectrum. All of those that fall into the middle could be labeled as undetermined, as we cannot statistically distinguish them from the two extreme groups. This suggests that if we were to try to create groups so that all individuals in one group were statistically distinguishable from all members of another group, the groups we create would be composed of few people. This is only one way to create groups while accounting for statistical uncertainty at the individual level. Alternative methods, such as considering pairwise probabilities that one individual's likelihood of the outcome is higher than another's, may result in larger separable groups.   

The right panel of Figure \ref{fig:interval-examples} shows the average 95\% posterior credible length within each of the PSA's FTA risk groups as a function of the number of times each individual was previously observed in the dataset. As expected, the intervals tend to get smaller as  the number of observations grows. However, we find that the average interval length, even for individuals that have been observed several times remains large, and is largest for the people in the highest risk group.

\begin{figure}
    \centering
    \includegraphics[height=1.75in]{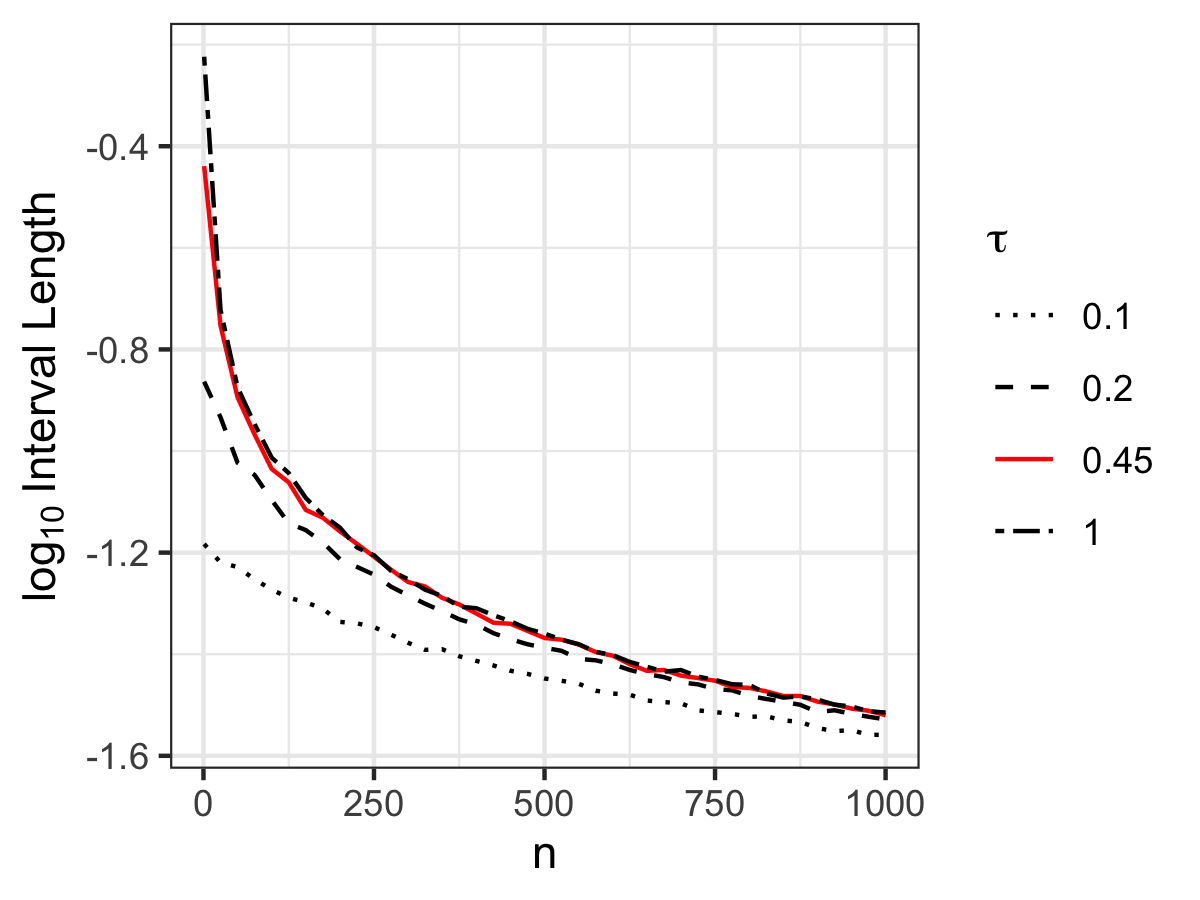}
    \includegraphics[height=1.75in]{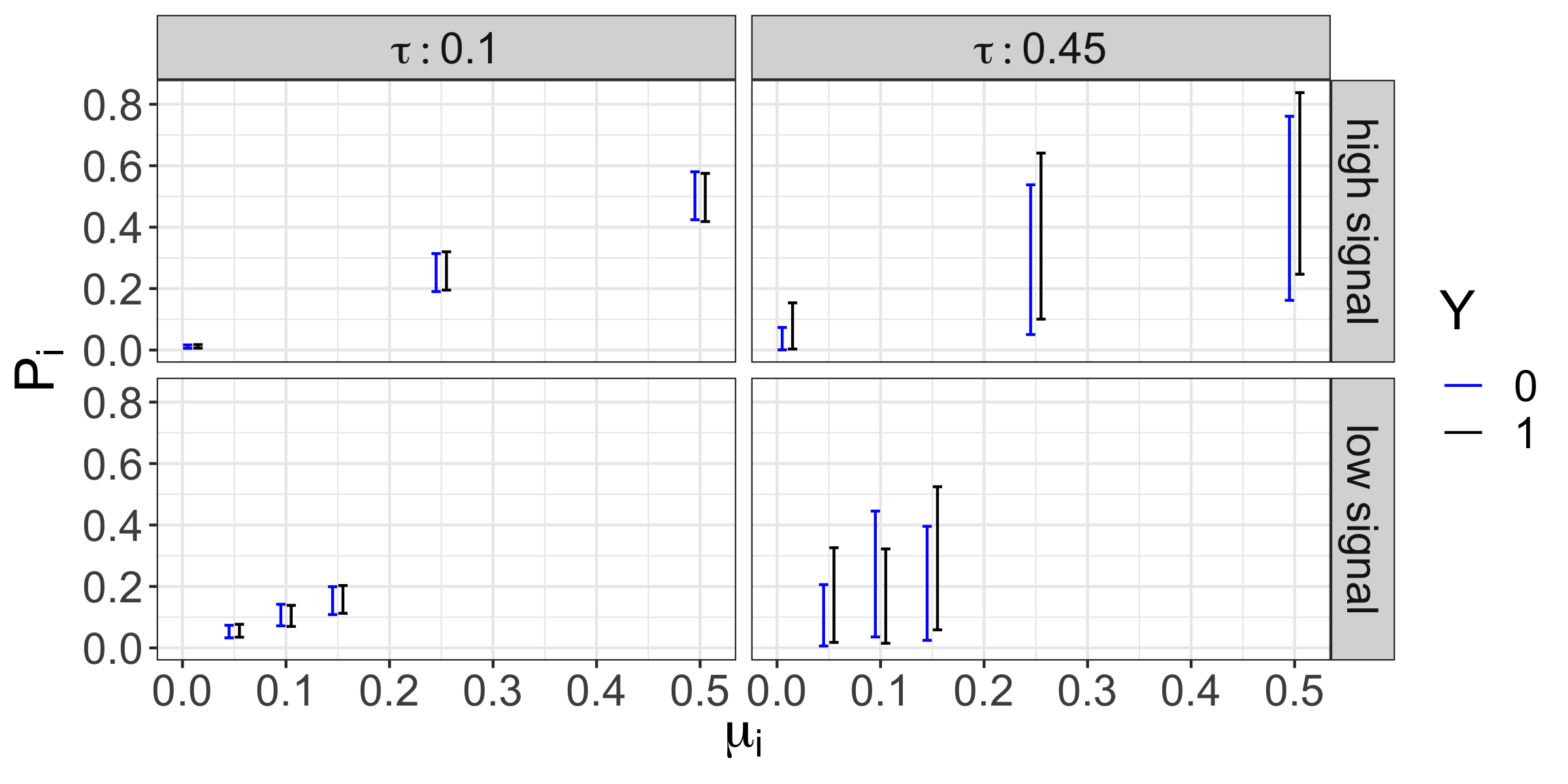}

    \caption{(left) Average length  of 95\% credible interval for $P_{ij}$ over 1000 replicates for each value of $n$. (right) 95\% posterior credible intervals for $P_{ij}$ for low/high signal and low/high noise scenarios. Intervals are colored by the value of the outcome variable.}
    \label{fig:simulation-plot}
\end{figure}

This raises the question of how many times an individual would need to be released and their outcome observed before one could say with a reasonable level of certainty what their individual $P_{ij}$ is. We consider an individual whose true probability falls right around the marginal rate in our data, and whose covariates perfectly predict their probability. For this individual, $0.20 = P_i = \Phi(X\beta)$, i.e. $\mu_i = 0$, though we don't assume $\mu_i$ to be known. We assume $\beta$ and $\tau$ are fixed. We simulate 1000 replicates of $Y_{ij} \stackrel{iid}{\sim} \Bern(0.2)$, for $j = 1, ..., n$ and $n = 1, ..., 1000$. For each value of $n$ and each of the thousand replicates, we calculate the posterior distribution of $P_{in}$ conditional on the $n$ observations of $Y_{ij}$. The length of the 95\% credible interval, averaged over replicates,  is shown in the left panel of Figure \ref{fig:simulation-plot} for several values of $\tau$. At the approximate value of $\tau$ estimated in our data (shown in red), posterior intervals for $P_{ij}$ remain fairly broad, even for unrealistically large values of $n$, the number of times the individual is released. At around $n=100$, for example, the average interval length drops below 0.10. However, if it were the case that $\tau$ were smaller, i.e. we learned that individuals do not deviate much from the value indicated by their covariates, reasonably small posterior intervals can be obtained with much more modest and realistic sample sizes. For example, after one observation when $\tau = 0.1$, posterior credible intervals average around length 0.06. Thus, it is not an inherent property of our model's structure that individuals have wide intervals--- were it the case that $\tau$ were smaller, even with few observations, we would have high posterior certainty about the range of values $P_{ij}$ could take.

This alone would not solve the problem of our inability to statistically distinguish individuals from one another. The right panel of Figure \ref{fig:simulation-plot} shows posterior intervals for $P_{ij}$ for ``low signal" and ``high signal" scenarios. In the low signal scenario, $\Phi(X\beta)$ takes three discrete values: 0.05, 0.15, and 0.25. In the high signal scenario,  $\Phi(X\beta)$ takes values 0.01, 0.25, and 0.50. We also consider a ``high noise" and a ``low noise" scenario for $\tau$: $\tau = 0.45$ as in our data and $\tau = 0.10$, respectively. 95\% posterior credible intervals are colored by the value of the outcome variable, $Y$. In the case where there is both high signal and small $\tau$, it is possible to create groups of individuals whose intervals do not overlap. For example, this scenario occurs for $\tau = 0.1$ with ``high signal", which correspond to areas of the parameter space not precluded by our priors or model structure. Reality just more closely resembles the low signal high noise scenario. 

\begin{figure}
    \centering
    \includegraphics[width=4in]{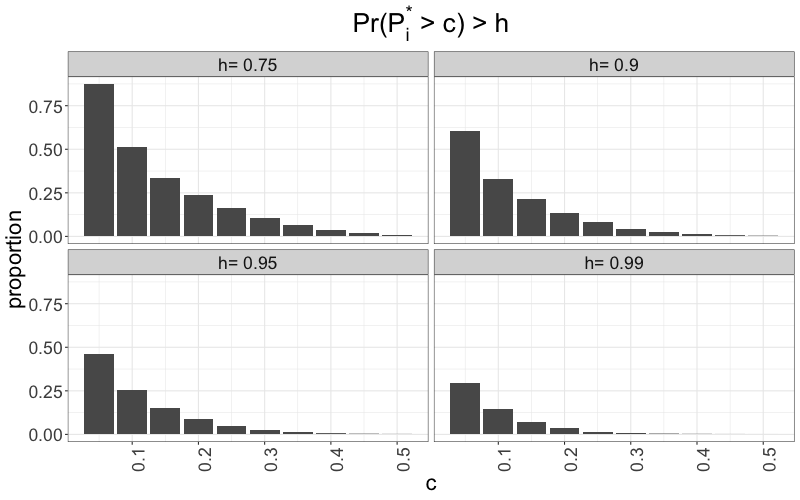}
    \caption{Proportion of the population in our data who would be flagged  under a decision-rule in which an individual is flagged if we can say certainty $h$ that their individual probability of FTA is at least $c$.}
    \label{fig:new-threshold}
\end{figure}

Finally, this analysis points to the possibility of creating risk groups not by thresholding the point estimate, but rather, by selecting the individuals for whom we have reached an acceptable level of statistical certainty that their probability of failure exceeds some threshold. 
For example, instead of flagging individuals whose point estimates exceed some policy-relevant threshold, $c$, one might instead flag all individuals for whom the posterior probability that $P_{ij}$ exceeds $c$ is at least $h$. For example, we might flag for additional review all people for whom we can say with probability $h \geq 0.95$ under our model that their likelihood of FTA on the next occasion exceeds c = 0.25. Figure \ref{fig:new-threshold} shows the proportion of individuals in this data who would be flagged for additional review for four values of certainty ($h$) and across a range of cutoff values $c$. 

If we select $h=0.99$, i.e. we demand to know with at least 99\% certainty that the individual's probability of failure is greater than $c$, then we find that for all but the smallest $c$s (0.05, 0.10), only a small proportion of individuals would be flagged. There is no occasion on which we would be able to say with 99\% certainty that FTA is more likely than not.  If we were to relax the certainty standard, and demand only that we know with 90\% certainty that an individual will fail to appear with probability at least $c$ on the next occasion, then for all thresholds greater than $0.4$, we would flag almost no one. Even at the least demanding level of certainty shown, only a tiny proportion of individuals could be flagged as being more likely than not to fail to appear.  

\section{Limitations}
One of the main assumptions of random effects models is  the distributional family of the random effects.  However, when replacing the Gaussian random effects distribution with a discrete mixture -- a very different form for the random effect distribution -- virtually none of our conclusions change (see the Appendix). This suggests this is not a major issue in our case. Chief among the remaining assumptions is surely the linear dependence on covariates. However, we note that predictive performance of risk assessment models used in criminal justice is similar, even when different predictive factors or more complex model structures are used \citep{desmarais2016performance, johndrow2019algorithm}. Moreover, generalized linear models are among the most common method used in building RAI's, and thus our specification is as close to the existing practice as possible while also including individual-level random effects. This parallel to current practice was intentional, as the results we present can only be attributed to accounting for individual-level variability as opposed to larger scale modeling differences. This leaves the assumption that individual-level random effects are constant across time as the main limitation of our analysis. While we believe it is worthwhile to try relaxing this assumption, given the few observations per person available, any additional flexibility in the model will only increase the level of uncertainty about individual probabilities. So, to the extent that one believes that our results are unrealistic due to the rigidity of our modeling assumptions, any additional complexity introduced into this model would likely only serve to increase the inferred level of variability across individuals. 

\section{Conclusion}
Ultimately, RAIs are used to assign labels to real human beings, and those labels affect the trajectory of their lives. This is a paper about statistical confidence in those labels. Though standard methods of evaluating statistical confidence have focused on model-level or group-level measures, here we focus on statistical confidence that the label applied to each individual is an accurate reflection of their individual-specific probability of the outcome. 

We find that, when individual propensity toward the outcome that is not explained by covariates is explicitly modeled, there is significant uncertainty about the probability of the outcome for most individuals. As a result, there is also significant uncertainty about the risk group to which most individuals belong: the between-individual variability swamps what little signal there is in the covariates. Perhaps more concerning, we find that even if we had an unrealistic number of observations per person--- if each person were arrested and released, say, 50 times--- the random effect distribution we have learned from the data would still result in substantial uncertainty about group membership and individual probabilities. As a consequence, it seems that the only way to create risk groups that result in high confidence about most individual group assignments is to put the vast majority of individuals in the same group. Since the real-world application of risk assessment is founded on different treatment for individuals assigned to different groups, allocating most individuals to one group would prevent actionable risk assessment. 

This analysis has uncovered an important difference between group-level and individual-level conclusions. While past discussion of group versus individual probabilities in this domain has been mostly theoretical due to an inability to meaningfully model individual probabilities using previously available data, our model moves the debate from theoretical to empirical. At the group-level, we can say with high certainty that the average rates within groups are different (for all but the highest two risk groups using the PSA risk groups). These sorts of results have been available in the past and, in fact, have been a standard component of model validations. Our model makes it possible to assess our uncertainty about an individual's probability of the outcome. In doing so, we can now also say that for any individual within the groups, we  do not in general have high confidence that their probability of the outcome is similar to that ascribed to them due to their inclusion in a certain risk group. Their probability of the outcome may be very different than that of the group to which they are assigned. 

Both notions of uncertainty are relevant. When evaluating policies at the population level one might be more concerned with the former. When faced with making decisions about an individual on the basis of the scores, the latter may be more relevant. Indeed, similar legal tensions between group- and individual-level inferences have arisen in other contexts, such as expert testimony \citep{faigman2014group}. How these notions of uncertainty ought to be weighed in determining how and whether RAIs are used is largely a matter of policy and law, not statistics. However, we do venture one policy recommendation: the fact that we typically have low confidence about the risk group assignment of any individual should be conveyed to decision makers, both at the level of decision-makers deciding whether an RAI is appropriate in their jurisdiction and those making pre-trial decisions.

\section{Acknowledgments}
This work was supported by a grant from Arnold Ventures. We thank the Kentucky Court of Justice Pre-Trial Services for providing the data used in this analysis as well as context and useful conversations. Useful comments on this manuscript were provided by Logan Koepke, Mingwei Hsu, Ira Globus-Harris, Karen Levy, and Sarah Riley. Karen Levy suggested the title.  

\bibliographystyle{rss}
\bibliography{references} 

\appendix

\section{Results for Discrete Mixture Random Effect Distribution} \label{sec:disc-mix-results}

\begin{figure}
    \centering
    \includegraphics[width=6in]{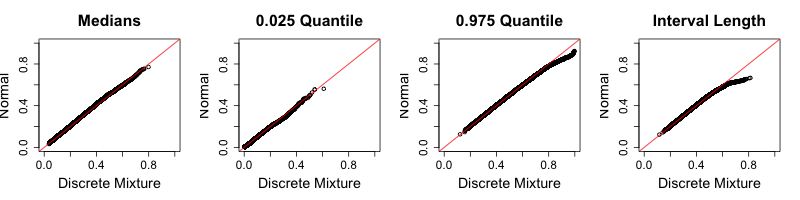}
    \caption{QQ-plots of the median and 0.025 and 0.975 quantiles of the distribution of $P_{ij}$ under the normal and discrete random effect distributions. We also show a QQ-plot of the interval length of the 95\% posterior credible interval.}
    \label{fig:model-comparison}
\end{figure}

We begin by showing a comparison of the distribution of $P_{ij}$ under the normal and discrete random effect distributions. We calculate several summary statistics of the distribution of $P_{ij}$ for each individual, including their posterior median, 2.5 percentile,  97.5 percentile, and the length of their 95\% posterior credible interval. Figure \ref{fig:model-comparison} shows QQ-plots of these summary statistics, showing the correspondence between the distributions of $P_{ij}$ under the two random effect distributions.  This shows that the discrete distribution tends to give a heavier right tail. Thus, results under this distribution paint a somewhat less optimistic view of the ability of the risk assessment model to disambiguate individuals. 

\begin{figure}
    \centering
    \includegraphics[width=5in]{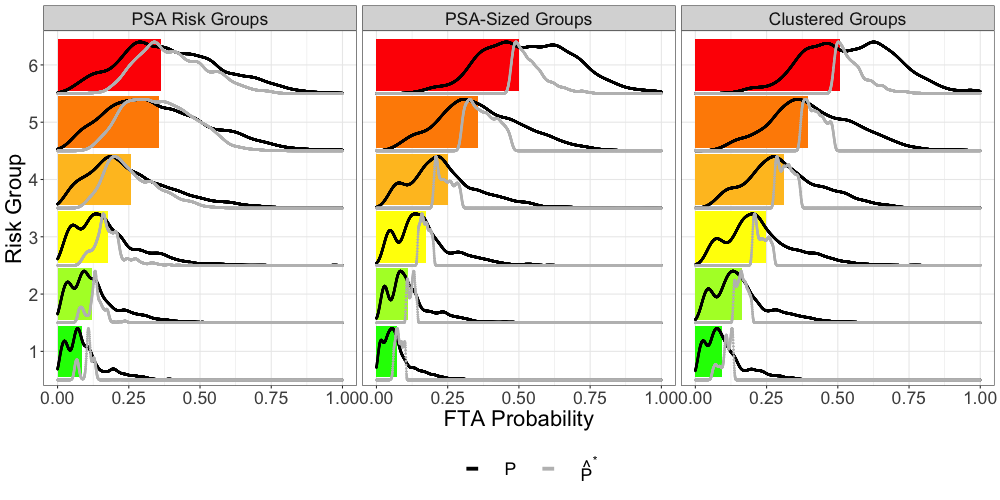}
    \caption{Estimated densities of $P_{ij}$ and $\hat{P}_{ij}$ under the discrete random effects distribution. }
    \label{fig:disc-mix-densities}
\end{figure}

\begin{figure}
    \centering
    \includegraphics[width = 5in]{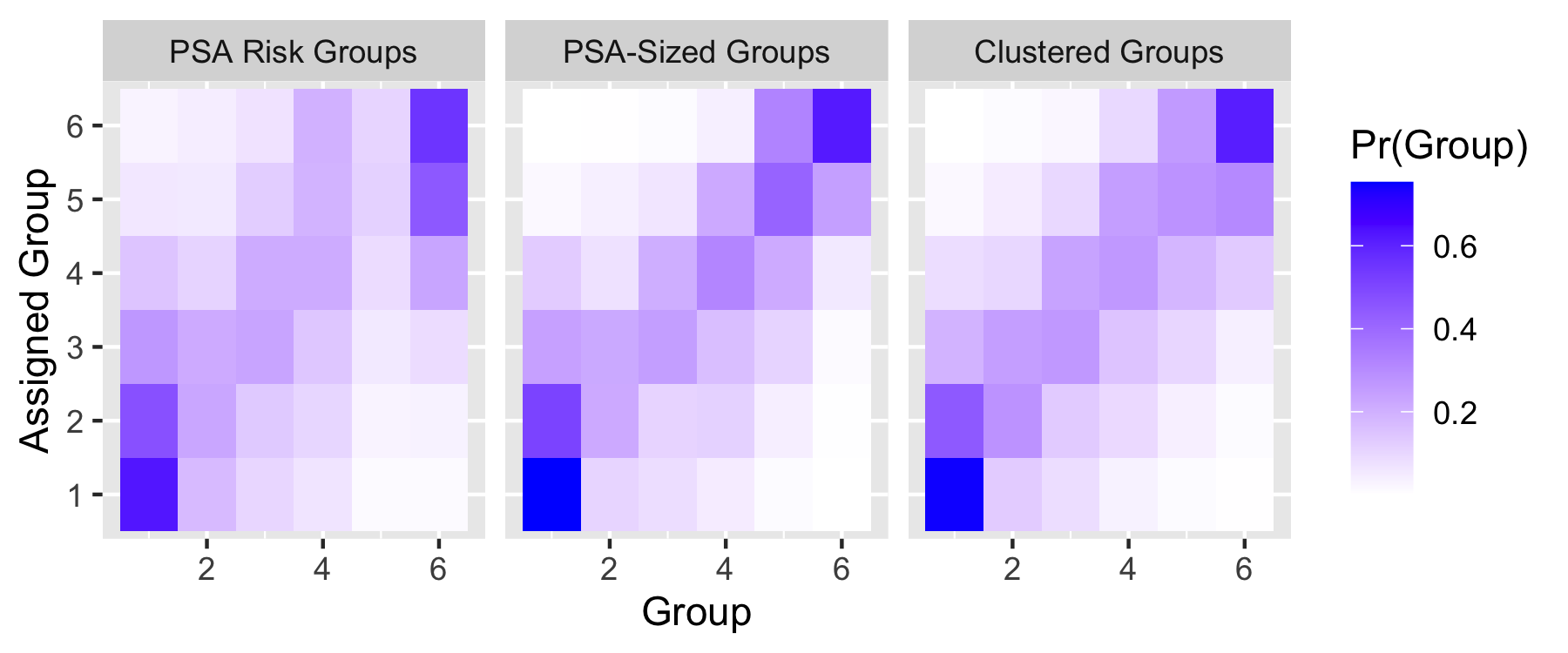}
    \caption{Posterior probability that $P_{ij}$ falls within the region of group $g$ (columns) for each actual group assignment ``Assigned Group" (rows). The color of the $kj$th cell corresponds to the posterior probability that $\tilde{g}_s(P_{ij}; c_s) = j$ given that $g_s(i) = k$ under the discrete random effect model.}
    \label{fig:disc-mix-wrong-group}
\end{figure}

Table \ref{tab:model-fit-disc-mix} shows summaries of the estimated rate of FTA by group under three different binning methods applied to the output of the discrete random effects model. This also shows good calibration. Figure \ref{fig:disc-mix-densities} shows the estimated densities of $P_{ij}$ and $\hat{P}_{ij}$ by risk group. Similar to the analysis with the normal random effects distribution, we find substantial individual-level heterogeneity about the group-wise mean. Finally, Figure \ref{fig:disc-mix-wrong-group} shows the posterior probability that one's $P_{ij}$ belongs in each bin, conditional on grouping by point estimates. This also shows the highest degree of ``correct" grouping for those in the lowest risk group. 

\begin{table}
\caption{The average posterior mean under the discrete random effects distribution of $\hat P^*_{ij}$, $P_{ij}$, and $Y_{ij}$ (the ``true", empirical rate) within risk groups under three different grouping strategies.} \label{tab:model-fit-disc-mix}
\centering
\begin{tabular}{l|lll|lll|lll}
\toprule
    & \multicolumn{3}{c}{PSA Risk Groups} \vline & \multicolumn{3}{c}{PSA-Sized Groups} \vline & \multicolumn{3}{c}{Clustered Groups} \\
\midrule
 & P* & P & rate & P* & P & rate & P* & P & rate \\ 
  \hline
1 & 0.10 & 0.10 & 0.08 (0.08,0.09) & 0.08 & 0.08 & 0.07 (0.06,0.08) & 0.08 & 0.08 & 0.07 (0.07,0.08) \\ 
  2 & 0.13 & 0.13 & 0.12 (0.12,0.13) & 0.12 & 0.12 & 0.11 (0.11,0.12) & 0.12 & 0.12 & 0.11 (0.1,0.11) \\ 
  3 & 0.18 & 0.18 & 0.18 (0.17,0.18) & 0.17 & 0.17 & 0.17 (0.17,0.18) & 0.15 & 0.15 & 0.15 (0.14,0.15) \\ 
  4 & 0.26 & 0.26 & 0.26 (0.25,0.27) & 0.24 & 0.24 & 0.25 (0.24,0.26) & 0.19 & 0.19 & 0.2 (0.2,0.21) \\ 
  5 & 0.36 & 0.36 & 0.36 (0.35,0.36) & 0.37 & 0.37 & 0.36 (0.35,0.36) & 0.26 & 0.26 & 0.27 (0.26,0.28) \\ 
  6 & 0.41 & 0.40 & 0.36 (0.35,0.38) & 0.54 & 0.54 & 0.5 (0.48,0.52) & 0.42 & 0.42 & 0.4 (0.39,0.41) \\ 
   \hline

\end{tabular}
\end{table}

\begin{figure}
    \centering
    \includegraphics[height = 1.55in]{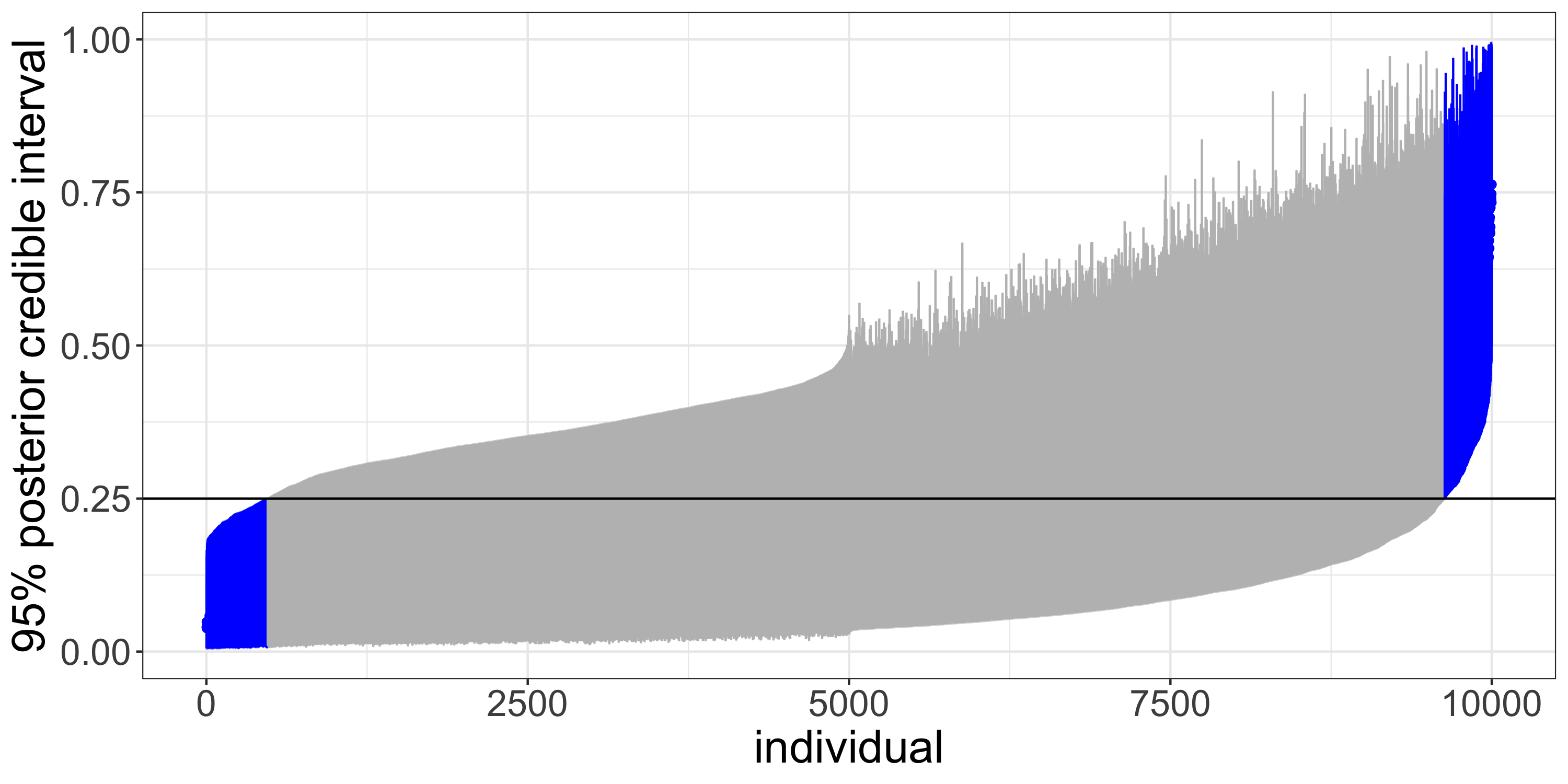}
    \includegraphics[height=1.55in]{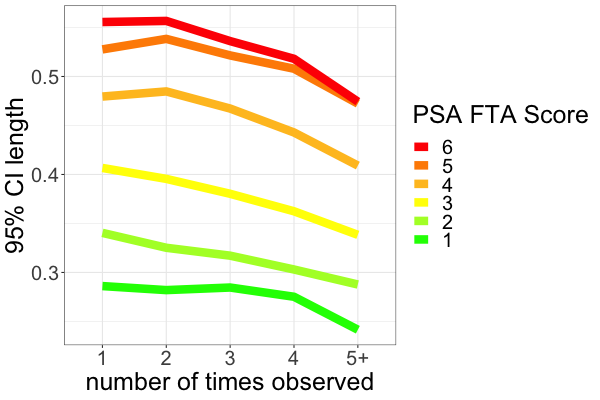}
    \caption{(left) Sample of 95\% posterior credible intervals under discrete random effect model. (right) Average interval length as a function of number of observations per person and risk group at last release.}
    \label{fig:intervals-disc-mix}
\end{figure}

\begin{figure}
    \centering
    \includegraphics[width=3in]{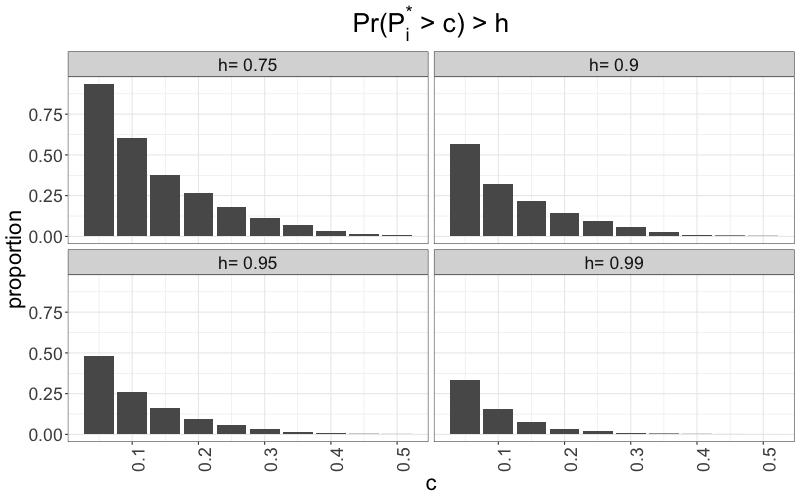}
    \caption{Proportion of population flagged for additional review under a decision rule accounting for uncertainty.}
    \label{fig:threshold-disc-mix}
\end{figure}

\newpage
\section{Supplementary Materials}

\subsection{Computation for the Discrete Random Effect Distribution}\label{sec:disc-mix-comp}

Now we consider the case where $f_\eta$ is a discrete distribution with $K$ classes, and we choose Gaussian priors on the atoms in the distribution. We take the approach implied by \cite{rousseau2011asymptotic}, which showed that in overfitted mixture models -- i.e. where the number of classes $K$ is greater than the true number of classes $K_0$ -- the posterior will concentrate around the true number of classes when a $\Dir(1/K,\ldots,1/K)$ prior is chosen on the mixture weights. The practical implication is that one can fit a finite mixture model with $K$ classes, and if many of the classes are unoccupied \emph{a posteriori}, then the selected value of $K$ can safely be considered large enough. This approach is a convenient alternative to specifying a prior on $K$ that nonetheless enjoys similar theoretical properties and leads to simpler computation. We choose $K=30$ and estimate the following model
\be
y_{ij} \mid z_i = k &\sim \Bern (\Phi(\theta_k + x_{ij} \beta) ) \\
\theta_k &\stackrel{ind}{\sim}  N(0,1), \quad \beta \sim N(0,9 I) \\
z_i &\sim \Cat(\nu), \quad \nu \sim \Dir(1/K,\ldots,1/K).
\ee
Let $Q$ be the number of occupied components in the mixture model. We found that with $K=30$, the posterior probability $\bb P[Q > 20] < 0.001$, and thus conclude that $K=30$ is sufficiently large. For computation, we again use a blocked data augmentation Gibbs sampler. However, the updates differ somewhat because of the different specification for $f_\eta$. Our Gibbs sampler has the following update rule.
\begin{enumerate}
    \item Update $z_i$ marginal of $\omega$ from $\Cat(\zeta_i)$ with $\zeta_i = (\zeta_i^{(1)},\zeta_i^{(2)},\ldots,\zeta_i^{(K)})$, where
    \be
    \log(p(\zeta^{(k)}_{i})) \propto \log(\nu_k) + \sum_{j=1}^{n_i} y_{ij} \log(\Phi(\theta_k + x_{ij} \beta)) + (1-y_{ij})\log(\Phi(-\theta_k-x_{ij}\beta))
    \ee

    \item Update $\omega$ given everything else, all conditionally independent, from
    \be
    \omega_{ij} \sim N_{y_{ij}}(\theta_{z_i} + x_{ij} \beta,1)
    \ee
    where as before $N_y$ is a normal truncated to $(0,\infty)$ if $y$ is positive and truncated to $(-\infty,0]$ if $y$ is nonpositive.
    
    \item Update $\beta$ given everything else from
    \be
    \beta &\sim N(\alpha_\beta,S_\beta) \\
    S_\beta &= (X'X + (1/9) I)^{-1},\quad \alpha_\beta = S_\beta X'(\omega - \theta).
    \ee
    where $\theta$ is the vector that has $\theta_{z_i}$ for all entries corresponding to person $i$
    
    \item Update $\theta$ given everything else, all conditionally independent, from
    \be
    \theta_k &\sim N\left( \alpha_k, s_k \right)   \\ 
    s_k &= \left( \sum_{i : z_i = k} n_i \right)^{-1}, \quad \alpha_k = s_k \sum_{i : z_i = k} \sum_{j=1}^{n_i} (\omega_{ij}- x_{ij} \beta).
    \ee
    
    \item Update $\nu$ given everything else from
    \begin{equation}
    \nu \sim \Dir\left( 1/K +\sum_{i: z_i = 1} n_i,\ldots,1/K + \sum_{i : z_i = K} n_i  \right).
    \end{equation}
\end{enumerate}
We run the algorithm for 80,000 iterations, discarding the first 40,000 iterations as burn-in.

\subsection{Derivation of full conditional for $\tau$}
To derive this, first write the data augmented model as
\be
\omega = \mu 1_N + \tau W \theta + X \beta + \epsilon
\ee
where $N = \sum_i n_i$ is the total number of observations of $y_{ij}$, $1_N$ is a column vector of $N$ ones, $\omega$ is the vectorization of $\omega_{ij}$, $\theta$ is the vectorization of $\theta_i$, and $W$ is the binary matrix that has a $1$ in its $k$th column when the corresponding row of $\omega$ corresponds to individual $i=k$ (i.e. it picks off the correct entry of $\theta$). Now since $\mu \sim N(0,c)$, we have $1_N \mu = \mu 1_N \sim N(0,c 1_N 1_N')$. So with $\mu$ marginalized out
\be
\omega \sim N(\tau W\theta + X\beta, I + c 1_N 1_N').
\ee
By the Woodbury matrix identity
\be
(I + c 1_N 1_N')^{-1} = I - 1_N (1_N' 1_N + c^{-1})^{-1} 1_N' = I - 1_N (N + c^{-1})^{-1} 1_N'.
\ee
So then after some simple algebra, we have that the log of the conditional density for $\tau \mid \theta, \omega, \beta$ (but marginal of $\mu$), up to a constant, is given by
\be
p(\tau \mid \theta, \omega,\beta) &\propto -\frac12 (\tau^2 \theta' (W' W -W' 1_N (N+c^{-1})^{-1} 1_N' W)  \theta \\
&\quad- \tau \theta' W' (I - 1_N (N + c^{-1})^{-1} 1_N') (\omega- X\beta)) - \frac12 \tau^2.
\ee
Now, noticing that $W'W = \diag(n_i)$ and $W' 1_N$ is the vectorization of $n_i$, we have, dropping the $-1/2$'s, putting $c = 9$,  and writing $\sum_{i,j}$ for $\sum_{i=1}^m \sum_{j=1}^{n_i}$
\be
p(\tau \mid \theta, \omega,\beta) &\propto \tau^2 \left\{ \left(1+ \sum_{i=1}^m \theta_i^2 n_i \right) - \left( \sum_{i=1}^m \theta_i n_i \right)^2 (N + 1/9)^{-1}  \right\} \\
&\quad- \tau \left[ \sum_{i=1}^m \sum_{i,j} \theta_i (\omega_{ij} - x_{ij} \beta) - \left(\sum_{i,j} n_i \theta_i\right)\left\{\sum_{i,j} (\omega_{ij} - x_{ij}\beta)\right\} (N + 1/9)^{-1} \right],
\ee
from which point it is easy to get \eqref{eq:TauFC}

\end{document}